\documentclass[prc,aps,preprint,nofootinbib,superscriptaddress,tightenlines]{revtex4}

\usepackage[english]{babel}
\usepackage[latin1]{inputenc}
\usepackage{graphicx}
\usepackage{amsmath}
\usepackage{amssymb}
\usepackage{color}
\usepackage{ulem}
\usepackage{bm}
\usepackage{xcolor}
\usepackage{hyperref}
\hypersetup{
    colorlinks,
    linkcolor={red!50!black},
    citecolor={blue!50!black},
    urlcolor={blue!80!black}
}

\usepackage{units}
\usepackage{subfigure}

\newcommand{\ket}[1]{\left|#1\right\rangle}

\newcommand{\spec}[2]{{}^{#1}\!#2}
\newcommand{\Iso}[2]{{}^{#2}\text{#1}}

\newcommand{\mN}{m_\text{N}}
\newcommand{\mc}{m_\text{c}}
\newcommand{\MNc}{M_\text{Nc}}
\newcommand{\muNc}{\mu_\text{Nc}}

\newcommand{\gs}{g_\sigma}
\newcommand{\gd}{g_\text{d}}
\newcommand{\Cl}[3]{\text{C}_{#1,#2}^{#3}}
\newcommand{\bs}[1]{\boldsymbol{#1}}
\newcommand{\Ecm}{E_\text{cm}}
\newcommand{\Ex}{E_\text{x}}
\newcommand{\Rc}{R_\text{c}}
\newcommand{\Rh}{R_\text{h}}

\newcommand{\OO}[1]{\mathcal{O}\left(#1\right)}

\newcommand{\tcm}{\theta_\text{cm}}
\newcommand{\Ed}{E_\text{d}}

\newcommand{\reaction}{$\Iso{Be}{10}(\text{d},\,\text{p})\Iso{Be}{11}$}

\newcommand{\etal}{\textit{et~al.}}

\def\d{\mathrm{d}}

\makeatletter
\newcommand{\raisemath}[1]{\mathpalette{\raisem@th{#1}}}
\newcommand{\raisem@th}[3]{\raisebox{#1}{$#2#3$}}
\makeatother

\begin{document}

\title{Neutron transfer reactions in halo effective field theory}
\author{M. Schmidt} \affiliation{Institut f\"ur Kernphysik, Technische Universit\"at Darmstadt, 64289 Darmstadt, Germany
}
\affiliation{Department of Physics and Astronomy, University of Tennessee, Knoxville, TN 37996, USA}
\author{L. Platter} \affiliation{Department of Physics and Astronomy, University of Tennessee, Knoxville, Tennessee 37996, USA}
\affiliation{Physics Division, Oak Ridge National Laboratory, Oak Ridge, Tennessee 37831, USA}
\author{H.-W. Hammer} \affiliation{Institut f\"ur Kernphysik, Technische Universit\"at Darmstadt, 64289 Darmstadt, Germany
}
\affiliation{ExtreMe Matter Institute EMMI, GSI Helmholtzzentrum f\"ur Schwerionenforschung GmbH,
64291 Darmstadt, Germany
}
\date{\today}

\begin{abstract}
Direct reaction experiments provide a powerful tool to probe the structure of neutron-rich nuclei like beryllium-11.
We use halo effective field theory to calculate the cross section of the deuteron-induced neutron transfer reaction \reaction.
The effective theory contains dynamical fields for the beryllium-10 core, the neutron, and the proton. In contrast, the deuteron and the beryllium-11 halo nucleus are generated dynamically from contact interactions using experimental and \textit{ab~initio} input. Breakup contributions are then included by construction. The reaction amplitude is constructed up to next-to-leading order in an expansion in the ratio of the length scales characterizing the core and the halo. 
The Coulomb repulsion between core and proton is treated perturbatively. Finally, we compare our results to cross-section data and other calculations.
\end{abstract}

\smallskip
\maketitle

\section{Introduction}
\label{Sec:Intro}
Nuclear processes such as capture and transfer reactions are one focus of ongoing research at existing and forthcoming experimental facilities with radioactive ion beams \cite{Fahlander:2013nfa}. However, the consistent theoretical description of such reactions in \textit{ab~initio} calculations poses significant challenges. Tremendous progress has been made for lighter systems in calculating elastic nucleus-nucleon scattering processes by combining the variational approach of the resonating group model and the no-core shell model in the no-core shell model with continuum \cite{Navratil:2016ycn}. However, for larger systems it remains a challenging task to calculate reactions in a controlled way and with reliable uncertainty estimates; see for example Refs.~\cite{Yoshida:2017pxa,Capel:2018kss,King:2018vzw,Nunes:2018erz,Lovell:2018bao}.

One alternative approach is to reduce the number of dynamical degrees of freedom. A process can then be described as an effective two- or three-body problem using a Lippmann-Schwinger or Faddeev equation. The remaining challenge is to model the interaction between the degrees of freedom appropriately. A reduction to the minimal degrees of freedom required to obtain a certain observable is frequently the starting point of an effective field theory (EFT) treatment of a system. EFTs can be applied if a system displays two disparate scales that can be combined to form a small expansion parameter. The large scale can for example be the excitation energy of a degree of freedom or a heavy state not included in the approach. EFT is the theory in which these high energy modes are integrated out.

Halo nuclei display such a separation of scales \cite{Zhukov:1993aw,Hansen:1995pu,Jonson:2004,Jensen:2004zz}. They consist of a tightly bound core with large excitation energy $\Ex$ and some weakly bound valence nucleons. The EFT that has been developed for these systems is called halo effective field theory (Halo EFT) \cite{Bertulani:2002sz, Bedaque:2003wa}. It treats the core as a fundamental degree of freedom, which is a valid approximation as long as energies smaller than $\Ex$ are considered. Halo EFT has been applied to a variety of processes including electromagnetic transitions and Coulomb dissociation of one-neutron halo nuclei. The formalism has been extended to one-proton and
two-neutron halo nuclei. For a recent review, see Ref.~\cite{Hammer:2017tjm}.

\begin{figure}
	\includegraphics[scale=1]{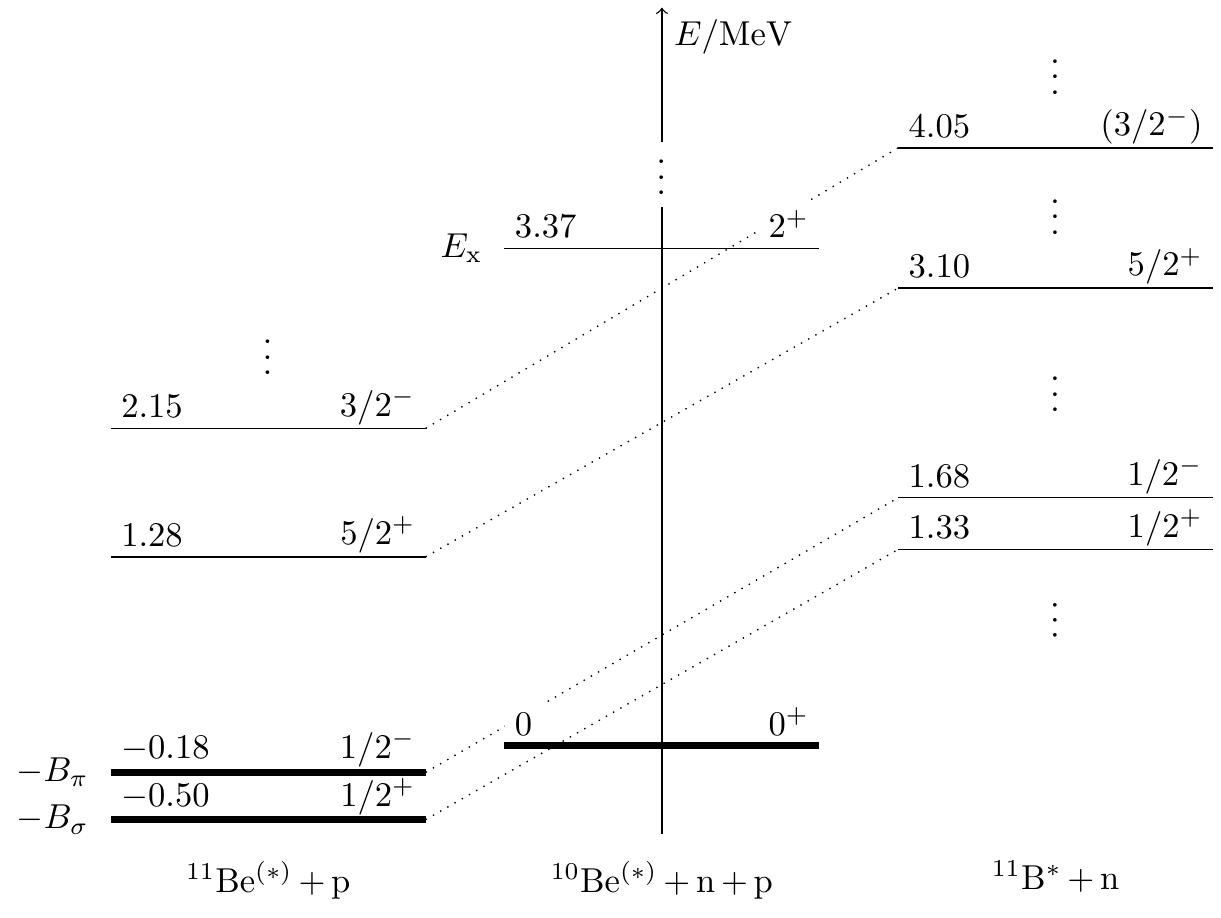}
	\caption{\label{Fig:Levels}Thresholds relative to $\Iso{Be}{10}$\,$+$\,n\,$+$\,p. The center column shows the ground and first excited state of beryllium-10. Bound and resonance states of the core-neutron (beryllium-11) and core-proton (boron-11) systems are depicted in the left and right columns, respectively. We only show $\Iso{B}{11}$ levels, which have been seen in the \mbox{$\Iso{Be}{10}(\text{p},\,\gamma)\Iso{B}{11}$} experiment of Ref.~\cite{Goosman:1970vn}. In this work, we explicitly include those states with thick lines.}
\end{figure}

In this work, we explore the potential of Halo EFT to describe the experimentally important process of a deuteron-induced transfer reaction. Such a calculation has not been carried out yet due to the challenging continuum structure of the reaction. As a test case, we consider \reaction. The effective three-body system is given by a $\Iso{Be}{10}$ core, a neutron, and a proton. The one-neutron halo nucleus $\Iso{Be}{11}$ represents a neutron-core state with a binding energy much smaller than the $2^+$ core excitation energy \mbox{$\Ex=\unit[3.37]{MeV}$}; see Fig.~\ref{Fig:Levels}. This intrinsic scale separation reflects itself also in the small core radius \mbox{$\Rc\sim \unit[2\text{-}3]{fm}$} and the large halo radius \mbox{$\Rh\sim \unit[7]{fm}$} \cite{Nortershauser:2008vp}. Exploiting these length scales, we construct the reaction cross section at leading order (LO) and next-to-leading order (NLO) in $\Rc/\Rh$. We find that dynamical core excitations and strong proton-core interactions can be neglected up to NLO. Deuteron and $\Iso{Be}{11}$ breakup contributions will be included automatically since Halo EFT contains all continuum states of the active degrees of freedom (core, proton, and neutron).

We expect that the Halo EFT expansion works best for center-of-mass energies $E$ well below \mbox{$\Ex=\unit[3.37]{MeV}$}; see Fig.~\ref{Fig:Levels}. However, in the absence of appropriate data, we compare our theory to data at \mbox{$E\geq \unit[7.78]{MeV}$}, measured by Schmitt \etal\ at Oak Ridge National Laboratory \cite{Schmitt:2012bt, Schmitt:2013uye}. In fact, previous works suggest that Halo EFT could still be appropriate for the lower experimental energies. For example, Deltuva \etal\ calculated the differential cross section in a Faddeev approach, using model interactions that reproduce elastic proton-core scattering data and optical potentials that account for loss channels\ \cite{Deltuva:2016his}. Their work suggests that core excitations barely influence the cross section for \mbox{$E\lesssim \unit[10]{MeV}$}. More recently, Yang and Capel~\cite{Yang:2018nzr} reanalyzed the reaction by combining the adiabatic distorted wave approximation reaction model with a Halo EFT description of $\Iso{Be}{11}$. They found out that, for the lower beam energies and forward angles, the reaction is purely peripheral. That is, it only depends on the asymptotic form of the $\Iso{Be}{11}$ wave function, while being independent of short-range details. Indeed, we will be able to describe data for the lower beam energies.

This manuscript is structured as follows. In Sec.~\ref{Sec:Lagrangian}, we present the EFT Lagrangian. Strong interactions among the core, neutron, and proton are described by contact forces and the Coulomb interaction follows from photon couplings. Section~\ref{Sec:2Body} explains how the two-body states $\Iso{Be}{11}$, $\Iso{Be}{11}^\ast$ and the deuteron emerge dynamically from the given interactions. We then turn to the three-body system in Sec.~\ref{Sec:3Body}. A Faddeev equation for the reaction will be constructed up to NLO in the $\Rc/\Rh$ expansion. Following work carried out for the three-nucleon sector \cite{Rupak:2001ci, Konig:2014ufa}, the Faddeev equation will include the dominant Coulomb contributions. After discussing results for the reaction cross section, we summarize our work and give an outlook in Sec.~\ref{Sec:SummaryOutlook}.

\clearpage

\section{EFT Lagrangian}
\label{Sec:Lagrangian}
The EFT Lagrangian $\mathcal{L}$ can be written as the sum 
\begin{equation}
  \label{Eq:Lagrangian-sum}
  \mathcal{L}=\mathcal{L}_1+\mathcal{L}_2+\mathcal{L}_3+\mathcal{L}_\gamma
\end{equation}
of one-, two-, and three-body interactions and a photon part. The one-body part reads
\begin{equation}
	\label{Eq:Lagrangian-1b}
	\mathcal{L}_1=
	n_\alpha^\dagger \left[
		i\partial_0+\frac{\bs{\nabla}^2}{2\mN}
	\right]n_\alpha
	+p_\alpha^\dagger \left[
		iD_0+\frac{\bs{D}^2}{2\mN}
	\right]p_\alpha
	+c^\dagger \left[
		iD_0+\frac{\bs{D}^2}{2\mc}
	\right]c\,.
\end{equation}
It introduces fields $n_\alpha$, $p_\alpha$
\mbox{($\alpha\in\{-1/2,\,+1/2\}$)} and $c$ for the neutron, proton, and $\Iso{Be}{10}$ core. They are treated as distinguishable particles. Sums over doubly appearing indices are
implicit. Masses are taken to be \mbox{$\mN\equiv \unit[938.918]{MeV}$} and \mbox{$\mc\equiv 10\,\mN$}.

The photon's kinetic and gauge fixing terms are given by
\begin{equation}
	\mathcal{L}_\gamma=-\frac{1}{4}F_{\mu\nu}F^{\mu\nu}-\frac{1}{2\xi} \left(
		\partial_\mu A^\mu -\eta_\mu \eta_\nu \partial^\nu A^\mu 
	\right)^2
\end{equation}
with timelike unit vector $\eta_\mu=(1,\,\bs{0})^T$. We only consider Coulomb photons, which induce a static potential. The covariant derivative \mbox{$D_\mu\equiv \partial_\mu + ieA_\mu \hat{Q}$} in Eq.~\eqref{Eq:Lagrangian-1b} with charge operator $\hat{Q}$ induces respective photon couplings $-ie\,Q_\text{p/c}$ with \mbox{$Q_\text{p}=1$} and \mbox{$Q_\text{c}=4$}. As done in
Ref.~\cite{Konig:2015aka}, we introduce a screened Coulomb photon
propagator
\begin{equation}
	iG_\gamma(\bs{p})\equiv i\left[
		\bs{p}^2+\lambda^2-i\epsilon
	\right]^{-1}.
\end{equation}
The artificial photon mass $\lambda$ has to be taken to zero at the end of each calculation.

The two-body part $\mathcal{L}_2$ involves the auxiliary fields
$\sigma_\alpha$ \mbox{($\alpha\in\{-1/2,\,+1/2\}$)} and $d_i$
\mbox{($i\in\{-1,\,0,\,+1\}$)} for the shallow bound states $\Iso{Be}{11}$ and deuteron, respectively. It reads
\begin{align}
	\nonumber
	\mathcal{L}_2
	=&\ \sigma_\alpha^\dagger \left[
		\Delta_\sigma^{(0)} -\left(
			i\partial_0+\frac{\bs{\nabla}^2}{2M_\text{Nc}}
		\right)
	\right]\sigma_\alpha
	-\gs \left[
		\sigma_\alpha^\dagger \left(n_\alpha c\right) + \text{H.c.}
	\right]
	\\
	\nonumber
	\label{Eq:Lagrangian-2b}
	&+d_i^\dagger \left[
		\Delta_\d^{(0)} -\left(
			i\partial_0+\frac{\bs{\nabla}^2}{4\mN}
		\right)
	\right]d_i
	-\gd\,\Cl{1/2\,\alpha}{1/2\,\beta}{1i'}
	\left[
		d_{i'}^\dagger (p_\alpha n_\beta) + \text{H.c.}
	\right]
	\\
	&+\mathcal{L}_{2,\Iso{Be}{11}^\ast}+\cdots
\end{align}
with \mbox{$\MNc\equiv \mN+\mc$} and a Clebsch-Gordan coefficient
$\Cl{s_1m_1}{s_2m_2}{s_3m_3}$. The expression ``H.c.'' denotes the
Hermitian conjugate. The regularization-dependent parameters
\mbox{$\Delta_a^{(0)},g_a\in\mathbb{R}$} \mbox{($a\in\{\sigma,\,\d\}$)} will be matched to experiment. Derivatives in Eq.~\eqref{Eq:Lagrangian-2b} induce range corrections at NLO. The part $\mathcal{L}_{2,\Iso{Be}{11}^\ast}$ accounts for further NLO contributions from the first excited state $\Iso{Be}{11}^\ast$. It is discussed in Appendix~\ref{App:Be11Exc}. Higher-order terms in the ellipses are negligible at NLO.

The three-body part $\mathcal{L}_3$ contains an $s$-wave deuteron-core interaction $C_0$ which will be used to renormalize the LO reaction amplitude. We write
\begin{equation}
	\label{Eq:Lagrangian-3b}
	\mathcal{L}_3
	=-\gd^2\,C_0\,(d_i c)^\dagger (d_i c) +\cdots\,.
\end{equation}

\section{Two-body states}
\label{Sec:2Body}
In this section, we show how $\Iso{Be}{11}$, $\Iso{Be}{11}^\ast$, and the deuteron emerge dynamically from contact interactions of the EFT Lagrangian. Our approach automatically takes care of two-body breakup, a crucial ingredient for the transfer reaction due to the small neutron separation energies of deuteron and $\Iso{Be}{11}$; see, for example, Refs.~\cite{Yilmaz:2001pi, Gomez-Ramos:2017ihe, DiPietro:2010zz}. Moreover, we explain the effective treatment of core excitation effects in the $\Iso{Be}{11}$ system.

\subsection{The beryllium-11 ground state}
\begin{figure}
	\centering
	\subfigure[]{
		\includegraphics[scale=1.2, trim=0 5 0 5, clip]{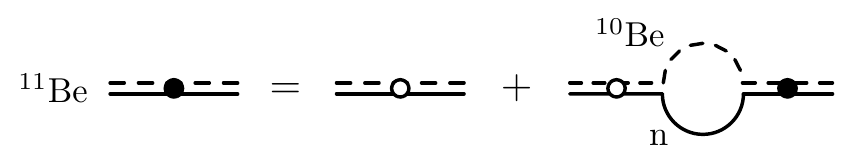}
	}
	\subfigure[]{
		\includegraphics[scale=1.2, trim=-2 5.5 0 5, clip]{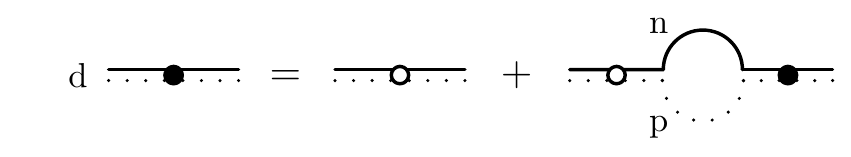}
	}
	\vspace{-0.2cm}
	\caption{\label{Fig:Dyson_sigma}(a) The full $\Iso{Be}{11}$ propagator $iG_\sigma$ (solid-dashed double line with filled circle) is obtained by coupling the bare one (empty circle) to the neutron-core continuum, represented by the self-energy loop (neutron: solid line; core: dashed line). (b) Similarly, the deuteron propagator $iG_\text{d}$ (solid-dotted double lines) couples to the neutron-proton continuum (proton: dotted line).
	}
\end{figure}

In Halo EFT, the $\Iso{Be}{11}$ ground state ($1/2^+$) is treated as a pure neutron-core $s$-wave state. Already at LO, its propagator, $iG_\sigma$, depicted as a solid-dashed double line in Fig.~\ref{Fig:Dyson_sigma} (a), contains iterations of the so-called neutron-core self-energy loop to all orders. This important quantity represents a summation over all neutron-core $s$-wave continuum states allowed by energy-momentum conservation. Thus, breakup contributions are automatically included.

As a consequence of the EFT's Galilean invariance, $iG_\sigma$ is a function of the center-of-mass energy \mbox{$\Ecm\equiv p^0-\bs{p}^2/(2\MNc)$} only, where
$p^\mu$ denotes the total four-momentum and $\MNc= \mN+\mc$ is the total mass. After resumming the self-energy loop, the propagator\footnote{The propagator is diagonal in spin
space. Respective factors $\delta^{\alpha\alpha'}$ will be omitted
in the following.} $iG_\sigma$ takes the well-known effective range expansion
form
\begin{equation}
	\label{Eq:EffRangeExp}
	iG_\sigma(\Ecm)= -i\,\gs^{-2}\,\frac{2\pi}{\muNc}
	\left[
		-a_\sigma^{-1} +\frac{r_\sigma}{2}k^2+\cdots -ik
	\right]^{-1},
\end{equation}
where $\muNc\equiv \mN \mc/(\mN+\mc)$ is the reduced mass and \mbox{$k\equiv i\,[-2\muNc(\Ecm+i\epsilon)]^{1/2}$} is the on-shell relative momentum \cite{Bethe:1949yr}. In the power divergence subtraction (PDS) scheme with mass scale
$\Lambda_\text{PDS}$ \cite{Kaplan:1998tg, Kaplan:1998we}, the
scattering length $a_\sigma$ and effective range
$r_\sigma$ are connected to the Lagrangian parameters $\Delta_\sigma^{(0)}$ and $\gs$ of Eq.~\eqref{Eq:Lagrangian-2b} by
\begin{align}
	\label{Eq:Matching_sigma}
	a_\sigma^{-1}&=\frac{2\pi}{\muNc}\frac{\Delta_\sigma^{(0)}(\Lambda_\text{PDS})}{\gs^2}+\Lambda_\text{PDS}\,,
	\\
	\label{Eq:Matching_sigma2}
  r_\sigma &= \frac{2\pi}{\muNc^2}\gs^{-2}\,.
\end{align}
The ellipses in Eq.~\eqref{Eq:EffRangeExp} denote higher-order terms. The unitary cut term $-ik$ is a manifestation of neutron-core continuum contributions.

The propagator has a pole at \mbox{$\Ecm=-B_\sigma$}, or equivalently at \mbox{$k=i\gamma_\sigma$}, where \mbox{$B_\sigma=\unit[0.50]{MeV}$} \cite{TUNL_Be11} and \mbox{$\gamma_\sigma\equiv (2\muNc B_\sigma)^{1/2}\approx\unit[29]{MeV}$} are the small binding energy and binding momentum. 
Thus, Eq.~\eqref{Eq:EffRangeExp} can be rearranged by writing
\begin{equation}
	\label{Eq:EffRangeExpPole}
	iG_\sigma(\Ecm)= 
	i\,\gs^{-2}\,\frac{2\pi}{\muNc}
	\left[
		\gamma_\sigma+ik
		-\frac{r_\sigma}{2}\left(
			k^2+\gamma_\sigma^2
		\right)
		+\cdots
	\right]^{-1},
\end{equation}
where we have expressed $a_\sigma^{-1}$ in terms of $\gamma_\sigma$ and $r_\sigma$.

Since the coupling $\gs$ is not an observable,
we eliminate it using redefined auxiliary fields
\mbox{$\tilde{\sigma}^{(\dagger)}_\alpha\equiv
\gs\,\sigma^{(\dagger)}_\alpha$}; see, for
example, Ref.~\cite{Griesshammer:2004pe}. Consequently, we have to multiply
$G_\sigma$
by $\gs^2$ and each (neutron-core)-$\Iso{Be}{11}$ vertex by
$\gs^{-1}$.

\subsubsection{Halo EFT counting and ANC}

In Halo EFT, all parameters
in Eqs.~\eqref{Eq:EffRangeExp}--\eqref{Eq:EffRangeExpPole} scale with certain powers of
the large halo radius \mbox{$\Rh\sim\unit[7]{fm}$} and the small core radius $\Rc\sim \unit[2\text{-}3]{fm}$. The latter represents the natural nuclear physics length scale
\cite{Hammer:2011ye}. We may estimate \mbox{$\Rc\sim (2\muNc \Ex)^{-1/2}\approx \unit[2.6]{fm}$} from the core excitation energy \mbox{$\Ex=\unit[3.37]{MeV}$}. The EFT expansion parameter is then given by \mbox{$\Rc/\Rh\sim 0.4$}.

As one of the first applications of Halo EFT to electromagnetic processes, Hammer and Phillips used data of the low-energy E1 strength of $\Iso{Be}{11}$ breakup, to
determine a value for $r_\sigma$ \cite{Hammer:2011ye}. Their result $\unit[2.7]{fm}$ scales like $\Rc$\,. In contrast, the binding momentum \mbox{$\gamma_\sigma\approx\unit[29]{MeV}$} is as small as $\Rh^{-1}\approx\unit[28]{MeV}$.
It follows that for low momenta \mbox{$k\sim \gamma_\sigma$}, the effective range term \mbox{$\sim \Rc\Rh^{-2}$} in Eq.~\eqref{Eq:EffRangeExpPole} is of NLO compared to \mbox{$\gamma_\sigma + ik\sim \Rh^{-1}$}.
Higher-order terms in the ellipses are of the order $\Rc^3\Rh^{-4}$ (N${}^3$LO) at most \cite{Hammer:2011ye}.

Once physics in the pole region is reproduced at a desired accuracy, it becomes obsolete to scale the $\Iso{Be}{11}$ ground-state wave function with a spectroscopic factor. Such scheme dependent quantities are not required in Halo EFT. Instead, Eq.~\eqref{Eq:EffRangeExpPole} yields an asymptotic normalization coefficient (ANC)
\begin{equation}
	\label{Eq:ANC}
	A_\sigma = \sqrt{
		\frac{
			2\gamma_\sigma}{
		1-\gamma_\sigma r_\sigma +\OO{\Rc^3/\Rh^3}
		}
	}
\end{equation}
for the radial wave function
\mbox{$u_\sigma(r)=A_\sigma \exp(-\gamma_\sigma r)$}, which is fully determined by low-energy observables \cite{Hammer:2011ye}.

Recently, Calci \etal\ were able to calculate the ANC using the no-core shell model with continuum \cite{Calci:2016dfb}. Their result \mbox{$A_\sigma=\unit[0.786]{fm^{-1/2}}$} was afterward confirmed by Yang and Capel in Ref.~\cite{Yang:2018nzr}, who extracted the value \mbox{$\unit[(0.785\pm 0.03)]{fm^{-1/2}}$} from the cross-section data of Ref.~\cite{Schmitt:2013uye}. The value was also confirmed in analyses of $\Iso{Be}{11}$ breakup at intermediate and high energies in Refs.~\cite{Capel:2018kss, Moschini:2018lwh}. We will use the ANC of Calci \etal\ as an input parameter at NLO. Equation~\eqref{Eq:ANC} can then be inverted to give a value for the effective range, which reads
\begin{equation}
	\label{Eq:rSigmaFromANC}
	r_\sigma \equiv \left(
		\gamma_\sigma^{-1}-\frac{2}{A_\sigma^2}
	\right)\left(1+\OO{\Rc^2/\Rh^2}\right)
	\approx \unit[3.5]{fm}\,.
\end{equation}
This value is larger than the one obtained by Hammer and Phillips in Ref.~\cite{Hammer:2011ye}. It will still be counted as $\Rc$, since \mbox{$\gamma_\sigma r_\sigma\approx 0.52$} differs by only \mbox{$0.12\lesssim (\Rc/\Rh)^2$} from \mbox{$\Rc/\Rh\sim 0.4$}.

\subsubsection{Propagator expansion}

From NLO, the propagator in Eq.~\eqref{Eq:EffRangeExpPole} exhibits spurious deep poles in addition to the physical one representing $\Iso{Be}{11}$ \cite{Ji:2011qg}. We solve this issue by expanding $iG_\sigma$ around \mbox{$k=i\gamma_\sigma$} in terms of $\Rc/\Rh$, yielding the series
\begin{align}
	\nonumber
	iG_\sigma(\Ecm)
	=&\ i\frac{2\pi}{\muNc}
	\left[
		\gamma_\sigma-\sqrt{-2\muNc (E_\text{cm}+i\epsilon)}
	\right]^{-1}
	\\
	\label{Eq:PoleExp_sigma}
	&\hspace{1cm}\times\left[1+
		\frac{r_\sigma}{2}
		\left(
			\gamma_\sigma+\sqrt{-2\muNc (E_\text{cm}+i\epsilon)}
		\right)
		+\OO{\Rc^2/\Rh^2}
	\right].
\end{align}
The residue of
$G_\sigma$ has an analog expansion and reads
\begin{equation}
	\label{Eq:Residue_sigma}
	Z_\sigma\equiv\Bigg[
		\left.
			\frac{\partial G_\sigma^{-1}}{\partial E_\text{cm}}
		\right|_{E_\text{cm}=-B_\sigma}
	\Bigg]^{-1}
	=\frac{2\pi}{\mu_\text{Nc}^2}\,\gamma_\sigma
	\left(
		1+\gamma_\sigma r_\sigma +\OO{\Rc^2/\Rh^2}
	\right).
\end{equation}

In Sec.~\ref{Sec:3Body}, $G_\sigma$ will enter the three-body Faddeev equation and $Z_\sigma$ is needed to normalize the reaction amplitude. At LO, we will truncate Eqs.~\eqref{Eq:PoleExp_sigma}--\eqref{Eq:Residue_sigma} after the
leading terms ``$1$,'' yielding expressions $G_\sigma^{(\text{LO})}$ and $Z_\sigma^{(\text{LO})}$. The NLO forms $G_\sigma^{(\text{NLO})}$ and $Z_\sigma^{(\text{NLO})}$ also include the terms linear in $r_\sigma$. We will follow Bedaque \etal\ by replacing \mbox{$G_\sigma^{(\text{LO})}\rightarrow G_\sigma^{(\text{NLO})}$} in the Faddeev kernel at NLO \cite{Bedaque:2002yg}. This straightforward technique is often referred to as ``partial resummation,'' because it induces specific amplitude terms proportional to $r_\sigma^n$, $n\geq 2$. In principle, such terms only occur at higher orders. However, for natural cutoffs, they are smaller then NLO terms and do not undermine the validity of the NLO calculation \cite{Ji:2011qg, Ji:2012nj}.

\subsubsection{Core excitation effects}

So far, we have treated $\Iso{Be}{11}$ as a pure \mbox{$1/2^+\otimes 0^+$} neutron-core state. However, in principle, it also couples to the \mbox{$1/2^+\otimes 2^+$} configuration of a neutron and a core excitation $\Iso{Be}{10}^\ast$ ($d$ wave). Note that this threshold resides far above the pole at an energy separation \mbox{$\Ex+B_\sigma\gg B_\sigma$}; see Fig.~\ref{Fig:Levels}. Close to the pole, $G_\sigma$ is insensitive to nonanalyticities of this remote channel.

Instead, it only receives residual modifications, which are automatically taken into account by renormalization onto low-energy observables $\gamma_\sigma$, $r_\sigma$, etc. Indeed, Deltuva \etal\ confirmed that dynamical core excitations within the $\Iso{Be}{11}$ bound state barely influence the reaction cross section \cite{Deltuva:2016his}. In other words, our \textit{effective} single-channel description readily contains all the relevant core excitation information in the pole regime. For illustration, we show in Appendix~\ref{App:CoreExc} that our approach is equivalent to a theory with an explicit $\Iso{Be}{10}^\ast$ field.

\subsection{The beryllium-11 excited state}

A second neutron-core state close to threshold is the first excited state $\Iso{Be}{11}^\ast$ ($1/2^-$). In Halo EFT, it is treated as a $p$-wave bound state \cite{Hammer:2011ye} with binding energy \mbox{$B_\pi=\unit[0.18]{MeV}$} \cite{TUNL_Be11}, or binding momentum \mbox{$\gamma_\pi\equiv(2\muNc B_\pi)^{1/2}\approx \unit[18]{MeV}$}. The Lagrangian part $\mathcal{L}_{2,\Iso{Be}{11}^\ast}$ is given in Appendix~\ref{App:Be11Exc}. As shown in Ref.~\cite{Bertulani:2002sz}, shallow $p$-wave states require the inclusion of at least two low-energy parameters. Close to the pole, we choose \mbox{$\gamma_\pi\sim \Rh^{-1}$} and the $p$-wave effective range \mbox{$r_\pi\sim \Rc^{-1}$}.
The propagator expansion then reads
\begin{equation}
	\label{Eq:PoleExp_pi}
	iG_\pi(\Ecm) = i\frac{6\pi}{\muNc}\frac{2}{-r_\pi}\left[
		\gamma_\pi^2+2\muNc(\Ecm+i\epsilon)
	\right]^{-1}\left(
		1+\OO{\Rc/\Rh}
	\right).
\end{equation}
Similarly to the ground state, $r_\pi$ can be obtained from the respective ANC $A_\pi$ \cite{Hammer:2011ye}. Taking the value \mbox{$A_\pi=\unit[0.129]{fm^{-1/2}}$} of Calci \etal\ \cite{Calci:2016dfb}, we find
\begin{equation}
	\label{Eq:rPiFromANC}
	r_\pi = -\frac{2\gamma_\pi^2}{A_\pi^2}\left(
		1+\OO{\Rc/\Rh}
	\right)\approx -\unit[0.95]{fm^{-1}}\,.
\end{equation}

In the transfer reaction \reaction, intermediate $\Iso{Be}{11}^\ast$ states represent NLO corrections to the reaction amplitude since \mbox{$G_\pi\propto\Rc<\Rh$}\,, and higher orders in Eq.~\eqref{Eq:PoleExp_pi} are at most of N${}^2$LO. For the moment, we neglect the excited state. It will be subject to the NLO discussion in Sec.~\ref{Sec:NLO}.

\subsection{The deuteron}

The deuteron is treated as an $s$-wave neutron-proton bound state with binding energy \mbox{$B_\d=\unit[2.22]{MeV}$} \cite{arXiv:nucl-th/9509032}. The product \mbox{$\gamma_\d r_\d\approx 0.40$} of the small binding momentum \mbox{$\gamma_\d\equiv (\mN B_\d)^{1/2}\approx\unit[46]{MeV}$} and the effective range \mbox{$r_\d=\unit[1.75]{fm}$}
\cite{arXiv:nucl-th/9509032} is as small as $\Rc/\Rh$. It follows that, up to NLO \mbox{($\sim \gamma_\d r_\d$)}, the deuteron propagator can be obtained in analogy to the one of $\Iso{Be}{11}$. In doing so, one also includes couplings of the deuteron propagator (solid-dotted double line) to the neutron-proton $s$-wave continuum; see Fig.~\ref{Fig:Dyson_sigma} (b).

After performing field redefinitions \mbox{$d^{(\dagger)}_i\rightarrow \tilde{d}^{(\dagger)}_i\equiv \gd\,d^{(\dagger)}_i$}, expressions for the propagator\footnote{The deuteron propagator is diagonal in spin space, i.e., it has to be multiplied by $\delta^{ii'}$ in diagrams.} $iG_\d$ around the pole, its residue $Z_\d$, and respective truncations can be obtained from
Eqs.~\eqref{Eq:PoleExp_sigma}--\eqref{Eq:Residue_sigma} by
replacing all subscripts ``$\sigma$'' by ``$\d$'', the total mass $\MNc$
by $2\mN$, and the reduced mass $\muNc$ by $\mN/2$. Relativistic effects and $s$-$d$ mixing are negligible up to NLO as
shown by Chen \etal\ \cite{Chen:1999tn}.

\subsection{Other partial wave channels}
Two-body interactions in partial waves different from the ones discussed above are negligible at NLO. For example, the $\spec{1}{S}_0$ virtual state of neutron-proton scattering enters the reaction \reaction\ at N${}^2$LO. Neutron-proton $p$-wave interactions enter at N${}^3$LO due to the lack of shallow states. Strong proton-core resonances shown in Fig.~\ref{Fig:Levels} would also enter at N${}^3$LO. Details on how to obtain these power counting classifications in Halo EFT will be given at the end of Sec.~\ref{Sec:NLO}.

Even though two-body interactions are restricted to channels with shallow states, the free (noninteracting) two-body continua will be taken care of in all partial wave channels; see below. These channels are described by plane waves up to NLO.

\section{Three-body system}
\label{Sec:3Body}
In this section, we derive an integral integration for the
reaction cross section from interactions of the Lagrangian
$\mathcal{L}$ up to NLO in the $\Rc/\Rh$ expansion. First, we show
which strong and Coulomb diagrams are induced by couplings of the Lagrangian $\mathcal{L}$. Second, we construct the LO transfer amplitude and present results for the LO cross section. At the end of the section, we discuss NLO corrections.

\subsection{Power counting and LO diagrams}
The transfer amplitude $T_{\sigma\d}$ connects the two states
\begin{equation}
	\ket{\sigma}\equiv\ \ket{\text{p}+\Iso{Be}{11}},\ 
	\ket{\d}\equiv\ \ket{\Iso{Be}{10}+\text{d}}
\end{equation}
through neutron exchanges and Coulomb diagrams. In EFT, these diagrams can be classified in a systematic power counting, which exploits the typical momentum scales of the system.

\subsubsection{Momentum scales}
The typical momentum scales of the three-body system are
given by the small binding momentum scale
\mbox{$\gamma\sim\gamma_\d\sim\gamma_\sigma\sim \Rh^{-1}$} and the inverse core
radius $\Rc^{-1}$. The largest subleading corrections in the strong
sector are suppressed by \mbox{$\gamma_\sigma r_\sigma\approx 0.52\sim\gamma_\d r_\d\approx 0.40$}; see above.

Coulomb diagrams additionally introduce the small ``Coulomb momentum''
\begin{equation}
	\label{Eq:CoulMom}
	p_\text{c}\equiv Q_\text{c}\,\alpha\,\muNc\approx \unit[25]{MeV}\lesssim \gamma\,,
\end{equation}
where \mbox{$\alpha\equiv e^2/(4\pi)\approx 1/137$} is the fine structure
constant. Moreover, Rupak and Kong pointed out that external momenta
$p$ have to be counted separately from $\gamma$ in the presence of
Coulomb photons \cite{Rupak:2001ci}. In this work, we calculate cross
sections for center-of-mass energies \mbox{$E\geq \unit[7.78]{MeV}$}. Thus, $p$ is of the
order \mbox{$p\sim (2\mN E)^{1/2}\geq \unit[120]{MeV}>\gamma$}. The two scales $p_\text{c}$ and $p$ form a second expansion parameter \mbox{$p_\text{c}/p< 0.2$}, which we will count like $(\Rc/\Rh)^2$.

\subsubsection{Strong interaction}

\begin{figure}
	\subfigure[]{
		\includegraphics[scale=1.2]{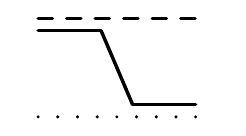}
	}
	\subfigure[]{
		\includegraphics[scale=1.2]{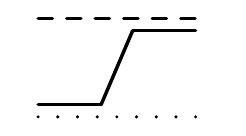}
	}
	\vspace{-0.3cm}
	\caption{\label{Fig:Vab}Neutron exchange diagrams (a)
          $-iV_{\sigma\d}$ and (b) $-iV_{\d\sigma}$. The different line types (single and double) are explained in Fig.~\ref{Fig:Dyson_sigma}. Both diagrams exhibit dynamical three-body intermediate states, coming into play via (a) $\Iso{Be}{11}$ and (b) deuteron breakup.}
\end{figure}

In Fig.~\ref{Fig:Vab}, we display the neutron exchange diagrams that
form the elementary building blocks of the strong interaction part of the transfer amplitude. We denote them by $-iV_{\sigma\d}^{Sm,1m'}$
and $-iV_{\d\sigma}^{1m,S'm'}$, where \mbox{$S,S'\in\{0,\,1\}$} and $m,m'$ represent total incoming and outgoing spins
and their projections, respectively.

Let $\bs{p}$ ($\bs{q}$) be the incoming (outgoing) relative momentum\footnote{In
  this work, relative momenta in the three-body center-of-mass system
  are defined as the momentum of the respective spectator particle. That is,
  they equal \mbox{$\bs{p}(\Iso{Be}{10})=-\bs{p}(\text{d})$} in
  $\ket{\d}$, or \mbox{$\bs{p}(\text{p})=-\bs{p}(\Iso{Be}{11})$} in
  $\ket{\sigma}$.} and $E$ the center-of-mass energy. We then find
\begin{align}
	\label{Eq:Vsd}
	V_{\sigma\d}^{Sm,1m'}\left(\bs{p},\,\bs{q};\,E\right)
	=&\ -\delta^{S1}\delta^{mm'}\,\mN\left[
		\bs{p}\cdot\bs{q} + p^2+\frac{1+y}{2}{q}^2-\mN(E+i\epsilon)
	\right]^{-1},
	\\
	\label{Eq:Vds}
	V_{\d\sigma}^{1m,S'm'}\left(\bs{p},\,\bs{q};\,E\right)
	=&\ V_{\sigma\d}^{S'm',1m}\left(\bs{q},\,\bs{p};\,E\right),
\end{align}
where \mbox{$y\equiv \mN/\mc$} is the mass ratio. Due to the $s$-wave nature of
the short-range interactions, only transitions between spin states
$S=S'=1$ with projections $m=m'$ are
possible.
In the following, we will refer to the functions in
Eqs.~\eqref{Eq:Vsd}--\eqref{Eq:Vds} as ``neutron exchange potentials''.

For neutron exchanges, we use the standard power counting of pionless EFT, which counts all momenta
formally like $\gamma\sim \Rh^{-1}$. Loops,
one-body propagators, and $s$-wave two-body propagators then count like
$\gamma^5/\mN$, $\mN/\gamma^2$, and $1/(\gamma\,\mN)$, respectively. It
follows that all neutron exchange iterations are of order
$\mN\Rh^2$ and have to be resummed at LO.

Recall that we include deuteron and $\Iso{Be}{11}$ breakup within the two-body-state propagators (double lines) to all orders by coupling them to continuum states as shown in Fig.~\ref{Fig:Dyson_sigma}. In three-body diagrams, further breakup contributions occur. For example, consider the diagram in Fig.~\ref{Fig:Vab} (a). The first\footnote{Time flows from left to right in our diagrams.} (upper) vertex in this diagram describes the breakup of the incoming $\Iso{Be}{11}$ bound state into a neutron-core pair. At this point, the initial $\ket{\text{p}+\Iso{Be}{11}}$ state evolves into an interacting $\ket{\text{p}+\text{n}+\Iso{Be}{10}}$ three-body state. Afterward, the exchanged neutron combines with the proton into a deuteron. Physically, the intermediate three-body state can be on shell since the center-of-mass energy $E$ is positive in the experiment by Schmitt \etal\ \cite{Schmitt:2012bt, Schmitt:2013uye}. Correspondingly, Eq.~\eqref{Eq:Vsd} exhibits poles for $E>0$.

\subsubsection{Coulomb contributions}
Next, we consider the Coulomb force, whose repulsion is expected to lower the reaction probability. In calculations, it
is usually included as a static two-body potential in addition to some
nuclear model interaction. In a strict EFT approach, however, Coulomb
diagrams can be analyzed in a systematic power counting, which exploits the system's momentum scales. This procedure reveals the relative importance of neutron exchange and Coulomb diagram interactions.

Photon couplings in $\mathcal{L}$ induce the diagrams $-i\Gamma_{ab}^{Sm,S'm'}$ \mbox{($a,b\in\{\d,\,\sigma\}$)} in
Fig.~\ref{Fig:CoulombDiagrams}. Their mathematical expressions are given in Appendix~\ref{App:CoulDiags}.
In the following, we analyze the diagrams using the Coulomb power counting suggested by Rupak and Kong \cite{Rupak:2001ci}.
\begin{description}
\item[Bubble diagrams]\itemsep0pt The one-loop diagrams (a) and (b) in
  Fig.~\ref{Fig:CoulombDiagrams} are proportional to
  the photon propagator ($\sim p^{-2}$) and to $p_\text{c}$; see Eq.~\eqref{Eq:CoulMom}.  All
  momenta in the loop (``bubble'') may be counted like
  $\gamma$.\,\footnote{This statement can be verified by analyzing the bubble diagrams in the limit of zero momentum transfer, where they are largest; see
    Appendix~\ref{App:CoulDiags}.} That is, we count one-body propagators
  like $\mN/\gamma^2$ and the loop integration by
  $\gamma^5/\mN$. The resulting scaling \mbox{$\mN\,\gamma\,p_\text{c}/(\gamma^2 p^2)$}
  suggests that bubble diagrams are small compared to neutron exchanges \mbox{($\sim\mN/\gamma^2$)} since \mbox{$p>\gamma\gtrsim p_\text{c}$}.

\item[Box diagrams]\itemsep0pt In the box diagrams of
  Figs.~\ref{Fig:CoulombDiagrams}(c) and \ref{Fig:CoulombDiagrams}(d), the photon is part of a
  loop. In this case, it is not straightforward to see if the
  corresponding integral is governed by powers of $p$ or
  $\gamma$. Since \mbox{$p>\gamma$} in our case, the safest option is to
  count the loop like $\mN/\gamma^3$. This scheme is in line with
  Ref.~\cite{Konig:2014ufa}. The overall scaling
  $\mN\,p_\text{c}/\gamma^3$ implies that box diagrams are of the same order as neutron
  exchanges since
  \mbox{$p_\text{c}\lesssim \gamma$}.
\end{description}

\begin{figure}
	\subfigure[]{
		\includegraphics[scale=1.2]{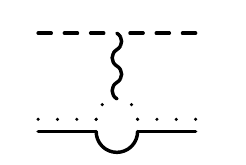}
	}
	\subfigure[]{
		\includegraphics[scale=1.2]{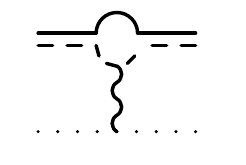}
	}
	\subfigure[]{
		\includegraphics[scale=1.2]{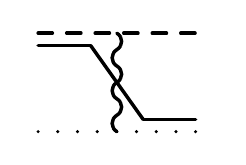}
	}
	\subfigure[]{
		\includegraphics[scale=1.2]{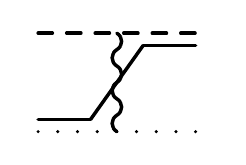}
	}
	\vspace{-0.3cm}
	\caption{\label{Fig:CoulombDiagrams}Coulomb diagrams entering the (a) $\ket{\d}$ channel, (b) $\ket{\sigma}$ channel, and [(c) and (d)] transfer channels. Curvy lines represent Coulomb photon propagators. In the LO calculation, each diagram is resummed to all orders; see Fig.~\ref{Fig:T_LO}.}
\end{figure}

In summary, the Rupak and Kong counting suggests that box
diagrams should be iterated at LO, while bubble diagrams
are subleading \mbox{($\sim\gamma\,p_\text{c}/p^2$)}. However,
one important feature of the bubble diagrams is not captured by the counting. Their photon propagators exhibit
infrared divergences at small momentum transfers in the limit of
vanishing photon mass; see
Eqs.~\eqref{Eq:Gamma_dd}--\eqref{Eq:Gamma_ss}. In principle, this enhancement could compensate for the discussed suppression. We account for this possibility by including the bubble
diagrams already in the LO
calculation, as was also done in Ref.~\cite{Konig:2014ufa}. We will then critically assess this choice by comparing the numerical influence of the box and bubble diagrams on the cross section.

Note that we only consider diagrams with one photon exchange
between two strong interactions. Corrections from two or more successive exchanges should be small since they involve further powers of the small Coulomb momentum $p_\text{c}$. In principle, they could be included by replacing each photon propagator with the full Coulomb $T$ matrix; see for example \cite{Konig:2014ufa}. We have checked that, for example, $-i\Gamma_{\d\d}$ would be modified by around \mbox{$\unit[20]{\%}\sim p_\text{c}/p\sim(\Rc/\Rh)^2$} in the on-shell limit. Such effects are neglected in this work.

\subsection{Transfer amplitude at LO}
\begin{figure}
	\includegraphics[scale=0.9]{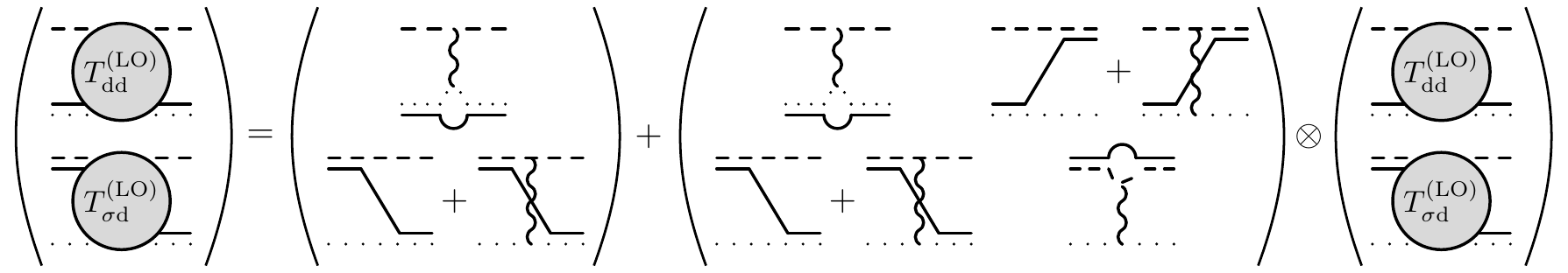}
	\caption{\label{Fig:T_LO}Transfer and elastic amplitude at LO. Loop
		integrals one the right-hand side contain LO propagators $G_{a}^{(\text{LO})}$ ($a\in\{\d,\,\sigma\}$) (drawn without circles). The
		three-body force $C_0(\Lambda)$ is omitted.}
\end{figure}

By iterating neutron exchanges, Coulomb bubble diagrams, and Coulomb box diagrams to all orders, we obtain the LO
transfer amplitude $T^{\text{(LO)}}_{\sigma\d}$. The corresponding Faddeev equation (without three-body force) is shown diagrammatically in Fig.~\ref{Fig:T_LO}. Loop integrals on the right-hand side ensure that all intermediate states allowed by energy-momentum conservation are taken care of.

\subsubsection{Partial wave channels}
It is beneficial for our purposes to perform a partial wave projection
onto the total angular momentum \mbox{$\bs{J}=\bs{L}+\bs{S}$} with
total spin $S$ and total orbital angular
momentum $L$. This procedure is explained in Appendix~\ref{App:PWProj}. The
respective neutron exchange potentials
\begin{align}
	\label{Eq:Vab_PW}
	V_{\sigma\d}^{\spec{2S+1}{L}_J,\spec{3}{L'}_J}\left(p,\,q;\,E\right)
	=&\
	\delta^{S1}\delta^{LL'}\,\frac{\mN}{pq}\,
	Q_L\left(
		-\frac{p^2+\frac{1+y}{2}q^2-\mN(E+i\epsilon)}{pq}
	\right),
	\\
	V_{\d\sigma}^{\spec{3}{L}_J,\spec{2S'+1}{L'}_J}\left(p,\,q;\,E\right)
	=&\ V_{\sigma\d}^{\spec{2S'+1}{L'}_J,\spec{3}{L}_J}\left(q,\,p;\,E\right),
\end{align}
depend on Legendre functions of the second kind,
\begin{equation}
	Q_L(x_0)\equiv -\frac{1}{2}\int_{-1}^1\!\d x\ 
	\frac{P_L(x)}{x-x_0}\,,
\end{equation}
in the convention of Ref.~\cite{AbramowitzStegun}.
Unfortunately, partial wave expressions of the Coulomb diagram interactions are impractically lengthy. Instead, we obtain them numerically by calculating
\begin{equation}
	\Gamma_{ab}^{\spec{2S+1}{L}_J,\spec{2S'+1}{L'}_J}
	\left(p,\,q;\,E\right)
	=\delta^{LL'}\,\frac{1}{2}
	\int_{-1}^1\!\d x\,P_L(x)\,\Gamma_{ab}^{S0,S'0}
	\left(\bs{p},\,\bs{q};\,E\right)\ 
	(a\in\{\d,\,\sigma\})
\end{equation}
with $x\equiv \bs{p}\cdot \bs{q}/(pq)$.

Cross sections will contain neutron exchange potentials and Coulomb contributions up to some $L_\text{max}$, at which results can be considered converged. It is worth noting that this approach does not only take care of higher partial waves between core-deuteron and proton-$\Iso{Be}{11}$. In fact, it automatically includes higher partial waves in each two-body sector (neutron-proton,\footnote{For example, when the proton-$\Iso{Be}{11}$ pair in Fig.~\ref{Fig:Vab} (a) is in $L=1$, then the intermediate three-body state has $L=1$ between the proton and an $l=0$ neutron-core pair. This configuration can be recoupled to $L=0$ between the core and an $l=1$ neutron-proton pair.} neutron-core, proton-core) due to breakup within the $-iV_{\sigma\d}$ and $-iV_{\d\sigma}$ diagrams; see Fig.~\ref{Fig:Vab}. Thus, the free two-particle continua (plane waves) are included up to $L_\text{max}$ although interactions are restricted to two-body channels with shallow states.

As indicated in Fig.~\ref{Fig:T_LO}, the LO elastic and
transfer amplitudes can be summarized into an amplitude vector $\vec{T}^{\text{(LO)}}$.
Due to the fact that the total spins \mbox{$S_\d=S_\sigma=1$} and orbital
angular momenta \mbox{$L_\d=L_\sigma\equiv L\in\{J-1,\,J,\,J+1\}$} are conserved at
LO, we identify a specific partial wave system by the superscript
``$[L,J]$.''  For incoming (outgoing) relative momenta $p$ ($p'$), we
finally obtain the scattering equations
\begin{align}
	\nonumber
	\vec{T}^{\text{(LO)}\,[L,J]}
	&\!\left(p,\,p';\,E\right)
	=- \underline{\underline{K}}^{\text{(LO)}\,[L,J]}
	\!\left(p,\,p';\,E\right) \cdot \vec{e}_1
	\\
	\label{Eq:LSEq_PW}
	&+4\pi\!\int\!\frac{\d q\,q^2}{(2\pi)^3}\ 
	\underline{\underline{K}}^{\text{(LO)}\,[L,J]}
	\!\left(p,\,q;\,E\right)\cdot
	\underline{\underline{\mathcal{G}}}^{\text{(LO)}}\left(q;\,E\right)\cdot 
	\vec{T}^{\text{(LO)}\,[L,J]}
	\!\left(q,\,p';\,E\right)
\end{align}
with LO amplitude vector, interaction matrix, and propagator matrix
\begin{align}
	\label{Eq:TVec_LO}
	\vec{T}^{\text{(LO)}\,[L,J]}
	\equiv&\ 
	\begin{pmatrix}
		T_{\d\d}^{\text{(LO)}}\\
		T_{\sigma\d}^{\text{(LO)}}
	\end{pmatrix}^{\spec{3}{{L}}_J,\spec{3}{{L}}_J}\,,
	\\
	\label{Eq:KMat_LO}
	\underline{\underline{K}}^{\text{(LO)}\,[L,J]}
	\equiv&\ \begin{pmatrix}
		\Gamma_{\d\d} & V_{\d\sigma}+\Gamma_{\d\sigma}\\
		V_{\sigma\d}+\Gamma_{\sigma\d} & \Gamma_{\sigma\sigma}
	\end{pmatrix}^{\spec{3}{{L}}_J,\spec{3}{{L}}_J}\,,
	\\
	\label{Eq:GMat_LO}
	\underline{\underline{\mathcal{G}}}^{\text{(LO)}}
	\equiv&\ \text{diag}\left(
		\mathcal{G}_\d^{\text{(LO)}},\,
		\mathcal{G}_\sigma^{\text{(LO)}}
	\right),
\end{align}
and \mbox{$\vec{e}_1\equiv (1,\,0)^T$} in channel space.
For convenience, we introduced the new functions
\begin{align}
	\label{Eq:NewPropFunctions}
	\mathcal{G}_a^{(\text{N${}^n$LO})}\left(q;\,E\right)
	\equiv&\ G_a^{(\text{N${}^n$LO})}\left(E-q^2/(2\mu_a)\right)\ 
	(a\in\{\d,\,\sigma\},\,n\in\mathbb{N}_0)\,,
\end{align}
where $\mu_\d\equiv 2\mN\,\mc/(2\mN+\mc)$ and
$\mu_\sigma\equiv (\mN+\mc)\mN/(2\mN+\mc)$.

The full transfer amplitude is given as a sum over the partial wave amplitudes and respective
projection operators as shown in Appendix~\ref{App:PWProj}. In all
calculations, we truncate the sum at some maximal orbital angular
momentum $L_\text{max}$ and increase this value toward
convergence. Similarly, whenever including Coulomb diagrams, we
decrease the photon mass \mbox{$\lambda\rightarrow 0$}. We find
that the cross section converges at \mbox{$L_\text{max}=12$} and
\mbox{$\lambda=\unit[0.1]{MeV}$}.

\subsubsection{Unphysical deep bound states}
To see if Eq.~\eqref{Eq:LSEq_PW} requires a three-body force for
renormalization, we have performed an asymptotic analysis for large
incoming and loop momenta
\mbox{$p,q\gg \gamma_\d,\gamma_\sigma,(\mN |E|)^{1/2}$} similar to
Ref.~\cite{Griesshammer:2005ga}.
In this limit, nucleon exchanges \mbox{($\sim q^{-2}$)}
dominate over Coulomb contributions \mbox{($\sim q^{-3}$)}
\cite{Konig:2014ufa}. Thus, we may neglect the
Coulomb force for the moment. It turns out that for \mbox{$L\geq 1$}, the potentials in Eq.~\eqref{Eq:LSEq_PW} fall off fast enough to produce unique amplitudes solutions. In the \mbox{$[L,J]=[0,1]$} system, however, that is not the case. Instead, the amplitudes approach a power law
behavior \mbox{$\sim p^{-1\pm is_0}$} with \mbox{$s_0=0.6357$}.
It follows that the system exhibits an Efimov effect, i.e., a geometric spectrum of three-body bound states at energies $E=-B_\d-B_3$
\cite{Efimov:1970zz, Braaten:2004rn, Naidon:2016dpf}. We note that \mbox{$\exp(\pi/s_0)\approx 140$} reproduces the universal scaling factor of three distiguishable particles with mass ratio \mbox{$y=0.1$} presented in Ref.~\cite{Braaten:2004rn}.

\begin{figure}
	\includegraphics[scale=1]{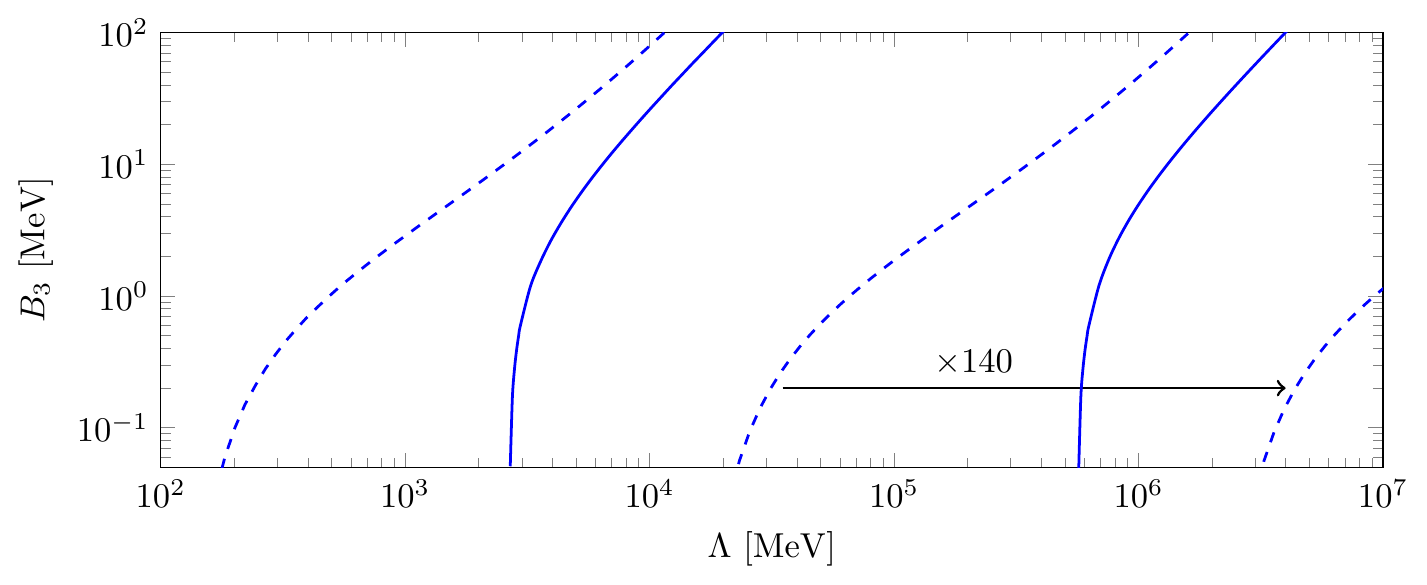}
	\vspace{-0.3cm}
	\caption{\label{Fig:Spectra}Unrenormalized three-body spectra
          at LO without (dashed lines) and with (solid lines) Coulomb
          diagrams ($\lambda=\unit[0.1]{MeV}$, converged) for
          various cutoffs $\Lambda$.}
\end{figure}

In the following, we equip Eq.~\eqref{Eq:LSEq_PW} with a momentum
cutoff \mbox{$\Lambda\gg \gamma_\d,\gamma_\sigma,(\mN |E|)^{1/2}$}. The
resulting spectrum is shown in Fig.~\ref{Fig:Spectra} as dashed
lines. Coulomb diagrams do not influence the large
momentum behavior of the system qualitatively. They only push the Efimov states to higher cutoffs (solid lines in
Fig.~\ref{Fig:Spectra}).
The system will be renormalized using the three-body coupling
$C_0(\Lambda)$ of Eq.~\eqref{Eq:Lagrangian-3b}. It enters the interaction
matrix of Eq.~\eqref{Eq:KMat_LO} as a constant $s$-wave potential like
\begin{equation}
	K_{\d\d}^{\text{(LO)}\,[0,1]}\rightarrow
	K_{\d\d}^{\text{(LO)}\,[0,1]}+C_0(\Lambda)\,.
\end{equation}
Note that the choice of this specific three-body force is not
unique. One could also introduce it in the transfer or the
$\ket{\sigma}$ elastic channel.

The quantum numbers of the Efimov states correspond to those of a
\mbox{$J^\pi=1^+$} level in boron-12. Experimentally, three such states
are
known~\cite{TUNL_B12}. In a deuteron-$\Iso{Be}{10}$ cluster
picture, their binding energies
\mbox{$B_3^\text{(phys)}\geq
\unit[5.77]{MeV}$} correspond to spatial separations
\mbox{$R_3=(2\mu_\d B_3)^{-1/2}\leq \unit[1.5]{fm}$} of the
deuteron-$\Iso{Be}{10}$ pair. Being of the order $\Rc$, they do
not reflect a separation of scales in the three-body sector. Thus,
the cluster picture is not justified and the Efimov states can be understood as artifacts of the short-range approach. However, although unphysical, they do not pose a problem as long
as they lie outside the EFT's region of applicability. Indeed, after
renormalization onto cross-section data, all three-body
states will occur at binding energies \mbox{$B_3>\unit[19]{MeV}$} and thus far away from the low-energy region; see Fig.~\ref{Fig:Renorm} (b).

\subsection{Cross section}
The differential cross section of the reaction \reaction\ at a deuteron beam energy
\begin{equation}
	\Ed= \frac{2\mN}{\mu_\d}\,\left(E+B_\d\right)
\end{equation}
can be obtained by multiplying the transfer amplitude by the residue factor $(Z_\sigma Z_\d)^{1/2}$ and evaluating it at on-shell relative
momenta,
\begin{equation}
	\bar{p}_a\equiv \sqrt{2\mu_a(E+B_a+i\epsilon)}\ 
	(a\in\{\d,\,\sigma\})\,.
\end{equation}

The cross section depends on the center-of-mass angle $\theta_\text{cm}$ with
\mbox{$\cos{\theta_\text{cm}}\equiv\hat{\bs{p}}(\text{d})\cdot
\hat{\bs{p}}(\text{p})$}.  In the $\ket{\d}$ channel, we set
the relative momentum to
\mbox{$\bar{\bs{p}}_\d\equiv -\bar{p}_\d\,\hat{\bs{p}}(\text{d})$} and in
the $\ket{\sigma}$ channel we take
\mbox{$\bar{\bs{p}}_\sigma\equiv \bar{p}_\sigma\,
\hat{\bs{p}}'(\text{p})$}.  The spin-averaged reaction cross section then reads
\begin{equation}
	\label{Eq:CrossSection}
	\left(\frac{\d\sigma}{\d\Omega}\right)\left(\theta_\text{cm};\,E\right)
	=\frac{1}{3}\,\sum_{m,S',m'}
	\frac{\mu_\d\mu_\sigma}{4\pi^2}\frac{\bar{p}_\sigma}{\bar{p}_\d}
	Z_d Z_\sigma \left|
		T_{\d\sigma}^{1m,S'm'}\left(\bar{\bs{p}}_\d,\,\bar{\bs{p}}_\sigma;\,E\right)
	\right|^2\,,
\end{equation}
where
\mbox{$|T_{\d\sigma}^{1m,S'm'}\left(\bs{p},\,\bs{p}';\,E\right)|^2
=|T_{\sigma\d}^{S'm',1m}\left(\bs{p}',\,\bs{p};\,E\right)|^2$}.

\begin{table}
	\caption{\label{Tab:InputParameters}EFT inputs for the calculation of the reaction cross section up to NLO in $\Rc/\Rh\sim 0.4$.}
	\begin{tabular}{p{0.2\textwidth}p{0.235\textwidth}p{0.235\textwidth}p{0.235\textwidth}}\hline\hline
		Order & deuteron & $\Iso{Be}{11}$ & $\Iso{Be}{11}^\ast$\\\hline
		LO [\,$\OO{1}$\,] & $B_\d=\unit[2.22]{MeV}$ \cite{arXiv:nucl-th/9509032} & $B_\sigma=\unit[0.50]{MeV}$ \cite{TUNL_Be11} & --\\[0.5em]
		NLO [\,$\OO{\Rc/\Rh}$\,] & $r_\d=\unit[1.75]{fm}$ \cite{arXiv:nucl-th/9509032} & 
		$A_\sigma=\unit[0.786]{fm^{-1/2}}$ \cite{Calci:2016dfb}
		& $B_\pi=\unit[0.18]{MeV}$ \cite{TUNL_Be11}\,,\\
		& & & $A_\pi=\unit[0.129]{fm^{-1/2}}$ \cite{Calci:2016dfb}
		\\\hline\hline
	\end{tabular}
\end{table}

Table~\ref{Tab:InputParameters} summarizes the input parameters needed for the calculation of the reaction cross section up to NLO in the $\Rc/\Rh$ expansion. At LO, only the binding energies $B_\d$ and $B_\sigma$ are required. At NLO, also the effective range $r_\d$, 
the ANC $A_\sigma$ of $\Iso{Be}{11}$, and the binding energy $B_\pi$ and ANC $A_\pi$
of $\Iso{Be}{11}^\ast$ enter.

\subsubsection{Coulomb suppression and improved LO system}
Our first goal is to critically assess the Coulomb power counting performed above. In particular, we would like to validate the proposed LO nature of the Coulomb force in general and of the bubble diagrams specifically, for the experimental energies used by Schmitt \etal~\cite{Schmitt:2012bt, Schmitt:2013uye}. Given the cutoff-dependence of the $L=0$ channel, we vary $\Lambda$ in the large range $\Lambda\in[300,\,1500]\,\unit{MeV}$ in each calculation. This procedure reveals the potential impact of the $s$-wave three-body force $C_0(\Lambda)$ on the LO reaction cross sections.

\begin{figure}
  \includegraphics[scale=1]{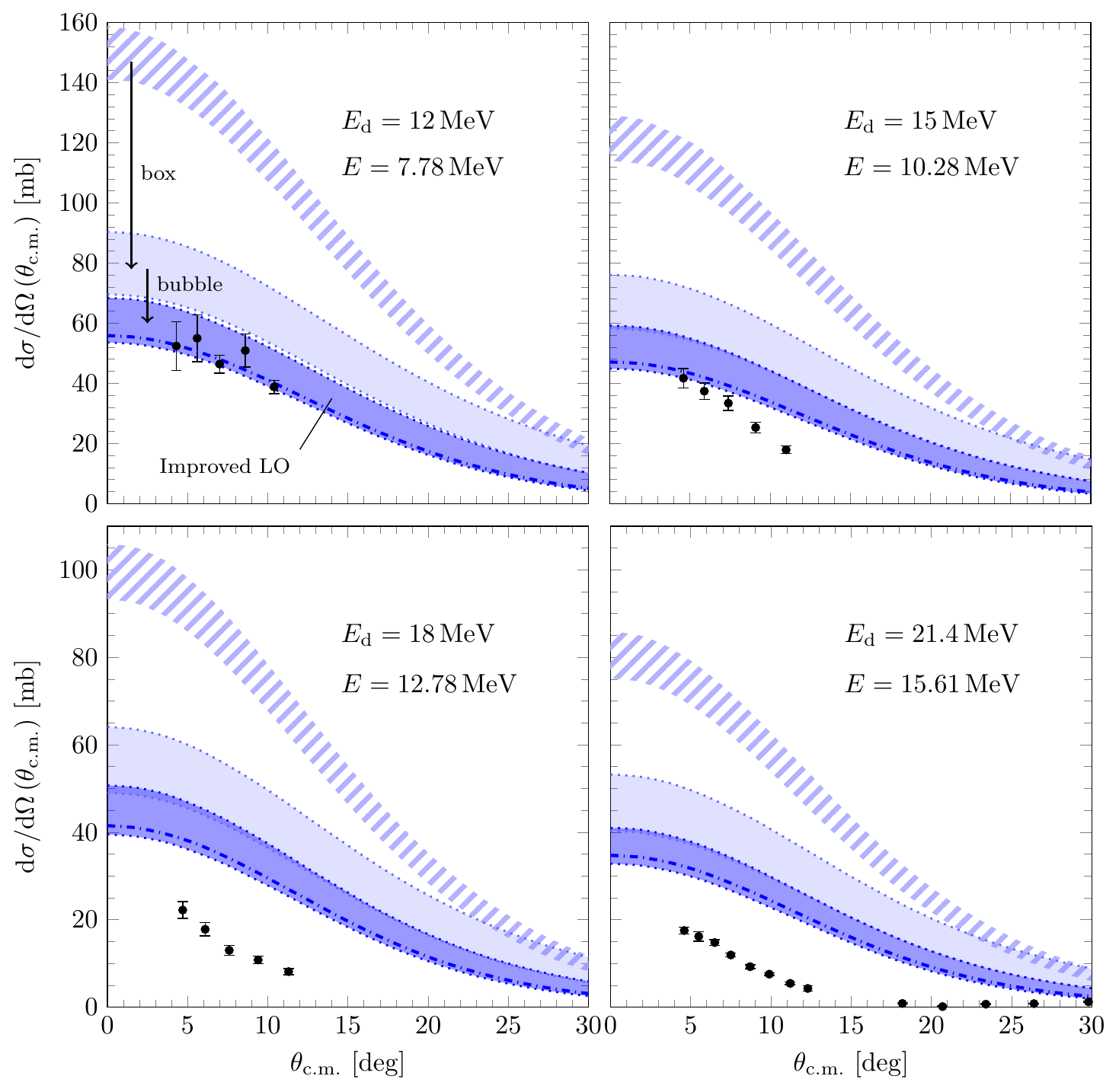}
  \vspace{-0.3cm}
  \caption{\label{Fig:CS_LO}LO cross section of \reaction\ as
    function of the center-of mass angle $\tcm$. For different
    deuteron energies $\Ed$ (lab frame), the results are compared to
    data (black points) from Ref.~\cite{Schmitt:2013uye}. All bands
    are due to cutoff variations
    \mbox{$\Lambda\in[300,\,1500]\,\unit{MeV}$}.
    Additional EFT uncertainties of order $\unit[40]{\%}$ due to neglected NLO contributions are omitted.
    Hatched bands exclude Coulomb
    contributions. Light (dark) bands enclosed by dotted lines
    include the Coulomb box (and bubble) diagrams. Dash-dotted
    curves represent a $\chi^2$ fit of the full equation system in
    Fig.~\ref{Fig:T_LO} onto the depicted \mbox{$\Ed=\unit[12]{MeV}$} data
    using the three-body force $C_0(\Lambda)$; see also
    Fig.~\ref{Fig:Renorm}. The fit is cutoff-independent for
    \mbox{$\Lambda\geq \unit[500]{MeV}$}. Each single curve is
    converged at \mbox{$L_\text{max}=12$} and
    \mbox{$\lambda=\unit[0.1]{MeV}$}.}
\end{figure}

In a first step, we neglect all Coulomb diagrams, which yields the uppermost bands (hatched) in Fig.~\ref{Fig:CS_LO}. Each curve is converged at percentage level for \mbox{$L_\text{max}=12$}. At all four deuteron beam energies \mbox{$E_{\d}\in\{12,\,15,\,18,\,21.4\}\,\unit{MeV}$} (lab frame), the bands lie high above the experimental data by Schmitt \etal\ \cite{Schmitt:2012bt, Schmitt:2013uye}. Apparently, the strong interaction alone does not produce enough repulsion between the scattering partners, even if $C_0(\Lambda)$ is included.

In order to understand the relative importance of the Coulomb box and bubble diagrams, we add them successively to the Faddeev equation. The light bands surrounded by dotted lines in Fig.~\ref{Fig:CS_LO} show that the box diagrams alone lower the cross sections drastically at all beam energies as expected. Indeed, it is important to include them at LO. Further repulsion comes from the bubble diagrams. Their inclusion yields the dark lowermost bands in Fig.~\ref{Fig:CS_LO}. Apparently, the influence of the bubble diagrams on the cross section is \mbox{$\lesssim \unit[40]{\%}$} smaller than the one of the box diagrams. Thus, it seems as if we have overestimated the enhancement due to the bubble diagrams' infrared divergences by one order in $\Rh/\Rc$. \textit{A posteriori}, the bubble diagrams are of NLO and could in principle be neglected at LO. The ``pure LO'' system then only contains neutron transfer and box diagrams.

Interestingly, however, the inclusion of the bubble diagrams as one specific NLO correction leads to a surprisingly good agreement with the cross-section data at lower beam energies and forward angles. Thus, choosing the ``improved LO'' system of Fig.~\ref{Fig:T_LO} significantly accelerates the EFT convergence. This statement will be verified later by including the remaining NLO corrections. Moreover, the improved LO system, unlike the pure one, can be renormalized onto data at \mbox{$\Ed=\unit[12]{MeV}$} since the respective band comprises all data points. We emphasize that none of the bands in Fig.~\eqref{Fig:CS_LO} includes the EFT uncertainties of $\pm\unit[40]{\%}$ at LO; see Fig.~\ref{Fig:CS_NLO} for comparison.

\subsubsection{Peripherality regions}

Although subleading in a strict sense, the bubble diagrams do not
introduce any new parameters like, for example, effective range coefficients. Thus, the improved LO system stays independent of short-range details. Cross sections are then only affected by the tail of the $\Iso{Be}{11}$ wave function, i.e., the reaction is purely ``peripheral''. Yang and Capel argued that such a description is sufficient to describe the reaction at lower beam energies and forward angles \cite{Yang:2018nzr}. Our results provide clear evidence for this claim since the improved LO band for \mbox{$\Ed=\unit[12]{MeV}$} perfectly describes the whole data region \mbox{($4.7^\circ\leq \tcm\leq 10.4^\circ$)}.

Moreover, according to Yang and Capel, the peripherality region
increases (decreases) in size for lower (higher) energies. Indeed, at \mbox{$\Ed=\unit[15]{MeV}$}, only forward scattering \mbox{($\tcm\leq 4.6^\circ$)} is captured by the improved LO band. Deviations at larger angles are of NLO size. At even higher energies \mbox{$\Ed\geq\unit[18]{MeV}$}, however, the
bands deviate from data by \mbox{$\unit[40\text{-}80]{\%}$}. We conclude that the reaction is indeed only peripheral at forward angles and low energies. For this reason our power counting may fail for energies \mbox{$\Ed>\unit[15]{MeV}$}.

Note, however, that Schmitt \etal\ identified their $\unit[18]{MeV}$ data set to be systematically smaller than the other three \cite{Schmitt:2013uye}. In particular, they extracted spectroscopic factors from all four data sets, of which the $\unit[18]{MeV}$ results were $\unit[25]{\%}$ smaller. Yang and Capel, who extracted the $\Iso{Be}{11}$ ANC from the data of Schmitt \etal, made a similar observation \cite{Yang:2018nzr}. Of all four data sets, only the $\unit[18]{MeV}$ set yielded an ANC $\unit[15]{\%}$ smaller than the prediction by Calci \etal\ \cite{Calci:2016dfb}. Thus, our calculation might be better at $\unit[18]{MeV}$ than suggested by Fig.~\ref{Fig:CS_LO}.

\subsubsection{Cutoff dependence and renormalizability}

Out of all components \mbox{$L\leq L_\text{max}=12$}, only the $L=0$ part is
cutoff-dependent. Due to this circumstance, the band widths in
Fig.~\ref{Fig:CS_LO} are only $\unit[20]{\%}$ the
size of the box diagram shift (LO). Such
contributions are negligible up to NLO. Thus, in principle, each curve within the filled bands
represents an LO result itself and renormalization is not required. Let us emphasize that the only inputs to our LO system are then given by the binding energies $B_\d$ and $B_\sigma$; see
Table~\ref{Tab:InputParameters}. At astrophysical energies, however,
the $L=0$ component is of much greater importance, leading to a much stronger cutoff dependence.

\begin{figure}
	\subfigure[]{
		\includegraphics[scale=1]{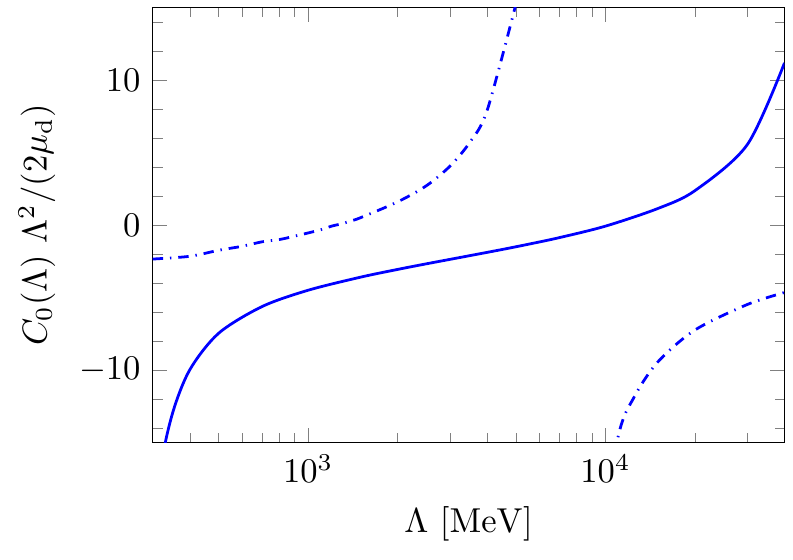}
	}\subfigure[]{
		\includegraphics[scale=1]{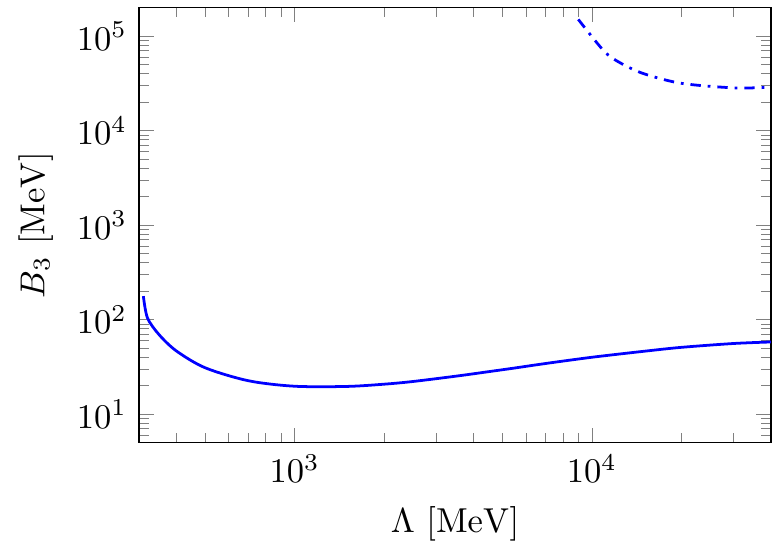}
	}
	\vspace{-0.5cm}
	\caption{\label{Fig:Renorm}Renormalization of the improved LO
          system depicted in Fig.~\ref{Fig:T_LO}. (a) The dot-dashed
          and solid curves are the two solutions of $C_0(\Lambda)$ for
          the $\chi^2$ fit to the $\Ed=\unit[12]{MeV}$ data set; see Fig.~\ref{Fig:CS_LO}. (b) Both solutions
          produce spectra outside the EFT regime, i.e., at binding
          energies \mbox{$B_3> \unit[19]{MeV}$} or \mbox{$B_3> \unit[28]{GeV}$},
          respectively.}
\end{figure}

We demonstrate the renormalizability of the improved LO system using the three-body force $C_0(\Lambda)$. For various cutoffs \mbox{$\Lambda\geq \unit[300]{MeV}$}, we adjust it in a
$\chi^2$ fit to the depicted $\Ed=\unit[12]{MeV}$ data set. This procedure yields the two solutions for
$C_0(\Lambda)$ shown in Fig.~\ref{Fig:Renorm} (a). Their fit values
\mbox{$\chi^2\approx 2.29$} (solid curve) and \mbox{$\chi^2\approx 2.23$}
(dot-dashed curve) are, within numerical uncertainties, equal in size
and respectively constant for \mbox{$\Lambda\geq \unit[500]{MeV}$}. For
illustration, we show fit results for \mbox{$\Lambda=\unit[500]{MeV}$} in
Fig.~\ref{Fig:CS_LO} as dot-dashed
curves. 
The first three-body state occurs at \mbox{$\Lambda\approx\unit[300]{MeV}$}
(or \mbox{$\Lambda\approx\unit[7]{GeV}$}); see Fig.~\ref{Fig:Renorm} (b). It
lies above \mbox{$B_3\approx \unit[19]{MeV}$} (or
\mbox{$B_3\approx \unit[28]{GeV}$}) and converges to even higher values as
\mbox{$\Lambda\rightarrow\infty$}.

\subsection{Corrections at NLO and beyond}
\label{Sec:NLO}
We now discuss NLO contributions to the reaction cross section in the $\Rc/\Rh$ expansion, stemming from range corrections in the two-body sectors and from the excited state $\Iso{Be}{11}^\ast$.

\subsubsection{Effective range corrections}
\begin{figure}
	\includegraphics[scale=1]{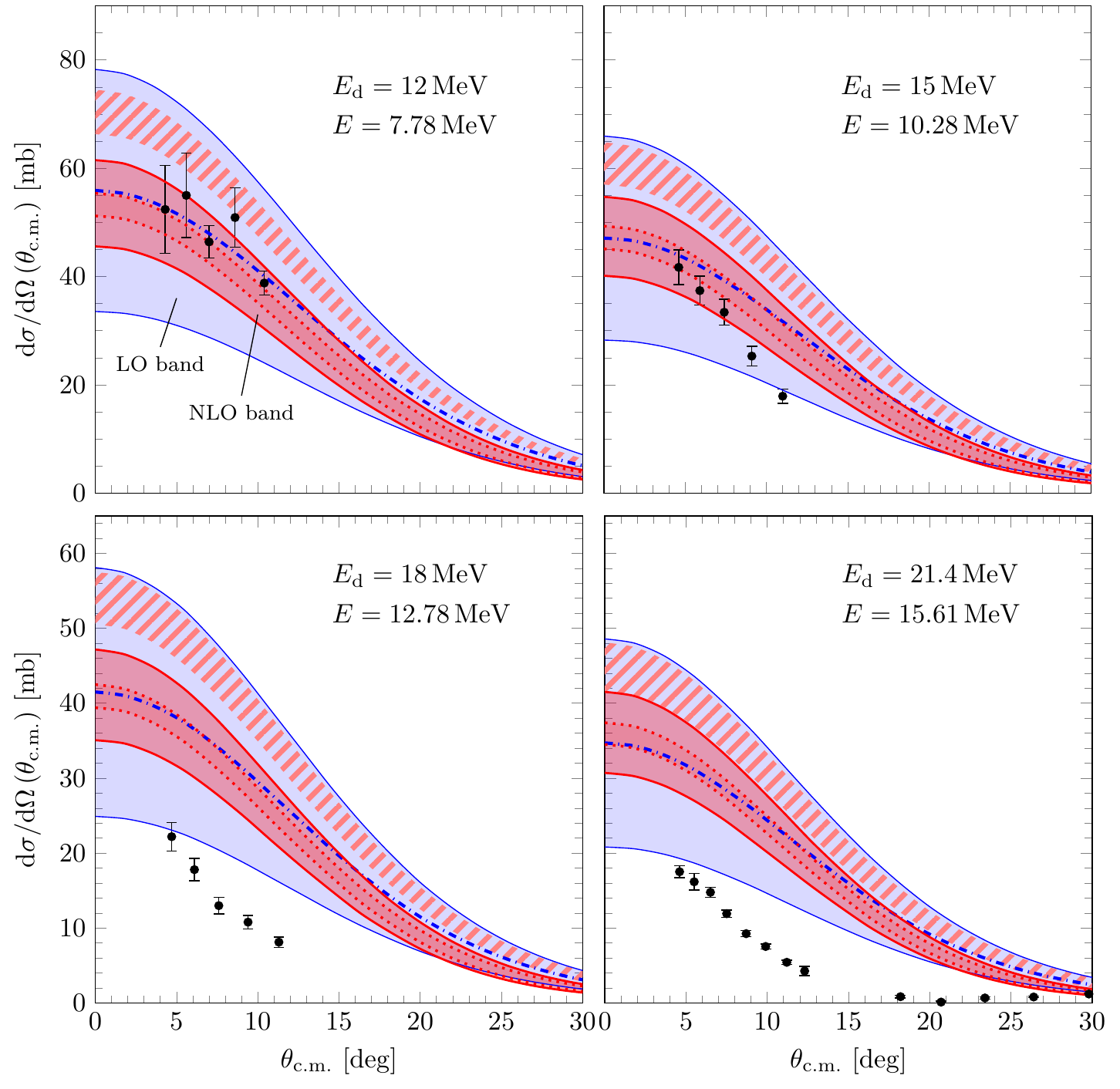}
	\vspace{-0.3cm}
	\caption{\label{Fig:CS_NLO}Cross section of \reaction\ up to
          NLO as function of the center-of-mass angle $\tcm$.
          Blue dot-dashed curves show the $\chi^2$ fits of the improved LO
          system. In contrast to Fig.~\ref{Fig:CS_LO}, we now vary these curves
          by $\unit[\pm 40]{\%}$ to mimic neglected NLO contributions. In doing so, we obtain the blue LO bands (enclosed by thin solid lines).
          The red hatched bands
          result from effective range corrections and the red
          bands enclosed by dotted lines also include corrections
          from the excited state $\Iso{Be}{11}^\ast$ at NLO.
          Both NLO calculations involve cutoff variations
          \mbox{$\Lambda\in[500,\,1500]\,\unit{MeV}$}. The final red NLO
          bands (enclosed by thick solid lines) represent
          $\unit[\pm 16]{\%}$ variations of the averaged NLO
          results due to neglected N${}^2$LO contributions.}
\end{figure}

A straightforward way to include effective range corrections in the deuteron and $\Iso{Be}{11}$ is to replace the LO propagators
$\mathcal{G}_a^{\text{(LO)}}$ by
$\mathcal{G}_a^{\text{(NLO)}}$ \mbox{($a\in\{\d,\,\sigma\}$)} in
Eq.~\eqref{Eq:LSEq_PW}
\cite{Bedaque:2002yg}.\footnote{Correspondingly, one has to use
  the residues $Z_a^{\text{(NLO)}}$ in the calculation of the cross
  section in Eq.~\eqref{Eq:CrossSection}.} This
approach reintroduces a cutoff dependence in the $L=0$ channel.
In principle, it could be cured by readjusting the three-body force $C_0(\Lambda)$ \cite{Hammer:2001gh}. In order to see the impact of the additional cutoff dependence, we
include effective range corrections in the renormalized improved LO
system for various
\mbox{$\Lambda\in[500,\,1500]\,\unit{MeV}$}\,.\,\footnote{Below \mbox{$\Lambda=\unit[500]{MeV}$}, the renormalized improved LO result is not yet converged. Note that the cutoff variation up to $\unit[1500]{MeV}$ is only used to estimate higher-order corrections. It does, however, not reveal the necessity of additional counter terms.} Figure~\ref{Fig:CS_NLO} shows that the resulting
red hatched bands lie well within the $\unit[\pm 40]{\%}$ LO
uncertainty bands (blue, enclosed by thin solid lines) of the improved LO estimates (blue dot-dashed curves). The band widths are comparably small, giving rise to a mild cutoff dependence.

It has to be mentioned that a small fraction of the band widths stems
from an unexpected cutoff dependence in the $L=1$ sector. It can be
understood as an artifact of the choice, not to perturb the amplitude
itself to first order in $\Rc/\Rh$, but the integration kernel. That
modifies the UV behavior of the partial wave amplitudes, leading to a
divergence in the $L=1$ sector. This divergence would not be present
in a strictly perturbative approach \cite{Griesshammer:2005ga}. Even
though desirable, such a more involved NLO treatment lies beyond the
scope of this work. In fact, we have checked that the influence of the cutoff on the $L=1$ amplitude is less than $\unit[2]{\%}$
over the range \mbox{$\Lambda\in[500,\,1500]\,\unit{MeV}$}. Thus, this
issue can be neglected at NLO.

\subsubsection{The beryllium-11 excited state}

The excited state $\Iso{Be}{11}^\ast$ introduces a third
channel \mbox{$\ket{\pi}\equiv\ket{\text{p}+\Iso{Be}{11}^\ast}$} to the
three-body system. It couples to $\ket{\d}$ via the diagrams $-iV_{\pi\d},\,-iV_{\d\pi}$ shown in Figs.~\ref{Fig:VabExc}(a) and \ref{Fig:VabExc}(b). Their mathematical forms and partial wave projections are given in Appendix \ref{App:Be11Exc}. We note that $\ket{\pi}$ only occurs as an intermediate state in the reaction. Thus, the NLO nature of $\Iso{Be}{11}^\ast$ follows from the propagator scaling \mbox{$G_\pi^{\text{(LO)}}\sim \Rc/(\gamma^2\,\mN)$}; see Sec.~\ref{Sec:2Body}. A typical contribution to
the reaction amplitude is given by Fig.~\ref{Fig:VabExc}(c). Again, we count all loop momenta like \mbox{$\gamma\sim \Rh^{-1}$}. The two (neutron-core)-$\Iso{Be}{11}^\ast$ vertices contribute a factor $\gamma^2$. The
overall scaling $\mN\,\Rc\Rh$ is then one order smaller
than the LO scaling $\mN\Rh^2$.

\begin{figure}
	\subfigure[]{
		\includegraphics[scale=1.2]{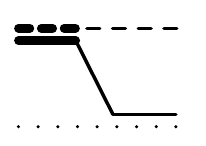}
	}
	\subfigure[]{
		\includegraphics[scale=1.2]{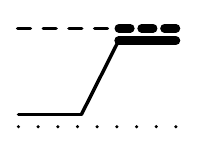}
	}
	\subfigure[]{
		\includegraphics[scale=1.2]{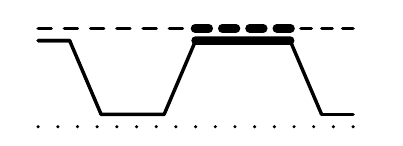}
	}
	\subfigure[]{
		\includegraphics[scale=1.2]{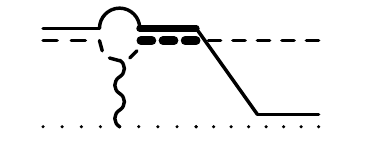}
	}
	\vspace{-0.2cm}
	\caption{\label{Fig:VabExc}The one-neutron exchange diagrams
          (a) $-iV_{\pi\d}$ and (b) $-iV_{\d\pi}$ induce NLO contributions from $\Iso{Be}{11}^\ast$ (thickened solid-dashed double line)
          to the transfer amplitude. The
          exemplary diagram in (c) contains a propagator
          $G_\pi^{\text{(LO)}}$ depicted as a thickened solid-dashed
          double line. In (d), the intermediate $\Iso{Be}{11}^\ast$ results from an E$1$ transition induced by the Coulomb field. This process is doubly suppressed by $G_\pi^{\text{(LO)}}$ and the photon propagator and can thus be neglected at NLO.}
\end{figure}

We complete the NLO system by inserting both effective range
corrections in $G_\d$ and $G_\sigma$, and the potentials
$V_{\pi\d},V_{\d\pi}$ into the integration kernel. The resulting
Faddeev equations are given in
Appendix~\ref{App:NLOEquations}. Similarly to the previous calculation, we vary
\mbox{$\Lambda\in[500,\,1500]\,\unit{MeV}$} and include the LO three-body
force $C_0(\Lambda)$. Figure~\ref{Fig:CS_NLO} shows that the results of
the previous calculation (hatched bands) get shifted back toward the improved LO results, ending up as red bands enclosed by dotted lines. Thus, the
influence of $\Iso{Be}{11}^\ast$ is indeed of NLO, in
agreement with our power counting. The remaining cutoff dependencies of the $L=0$ and $L=1$ sectors are negligible compared to
N${}^2$LO corrections ($\pm \unit[16]{\%}$, red uncertainty bands enclosed by
thick solid lines). Thus, no further renormalization is needed at NLO.

Recall that the NLO parameters \mbox{$r_\sigma=\unit[3.5]{fm}$} and \mbox{$r_\pi=\unit[-0.95]{fm^{-1}}$} were calculated in Eqs.~\eqref{Eq:rSigmaFromANC} and \eqref{Eq:rPiFromANC} from the ANCs of Calci \etal\ \cite{Calci:2016dfb}. Instead, one could directly use the Halo EFT values \mbox{$r_\sigma=\unit[2.7]{fm}$} and \mbox{$r_\pi=\unit[-0.66]{fm^{-1}}$} of Hammer and Phillips \cite{Hammer:2011ye}. The relative differences $\unit[30]{\%}$ and $\unit[40]{\%}$ are of size $\Rc/\Rh$ and should thus be negligible at NLO. We have checked that the final NLO bands would indeed only change by ca. $\unit[5]{\%}$. Thus, both choices for $r_\sigma,\,r_\pi$ are consistent with the proposed power counting.

In Ref.~\cite{Schmitt:2013uye}, the cross section for transfer to $\Iso{Be}{11}^\ast$ was also measured. In our theory, this quantity can in principle be calculated using the amplitudes $T_{\pi\d}$ in Eqs.~\eqref{Eq:TNLOVec1}-\eqref{Eq:TNLOVec2}. However, Yang and Capel found that this process is less peripheral than \reaction\ \cite{Yang:2018nzr}. For this reason, we expect that our low-energy power counting has to be modified in order to describe it. Indeed, naive application of the current scheme leads to an overestimation of the data.

\subsubsection{Higher-order interactions}

\begin{figure}
	\includegraphics[scale=1.2]{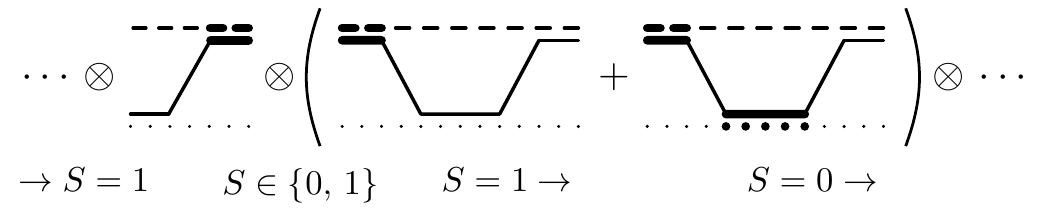}
	\vspace{-0.3cm}
	\caption{\label{Fig:VirtualState}The excited state $\Iso{Be}{11}^\ast$ allows transitions from total spin $S=1$ to $S=1$ ($\ket{\d}\rightarrow\ket{\pi}\rightarrow\ket{\d}$) or to $S=0$ ($\ket{\d}\rightarrow\ket{\pi}\rightarrow\ket{\Iso{Be}{10}+\text{np}(\spec{1}{S}_0)}$). The thickened solid-dotted double line represents the neutron-proton $\spec{1}{S}_0$ virtual state. Multiple transitions via $\ket{\pi}$ are negligible at NLO.}
\end{figure}

At higher orders in Halo EFT, additional interactions would enter the calculation. For example, the proton-neutron sector exhibits a shallow $\spec{1}{S}_0$ virtual state \cite{Kaplan:1998tg, Kaplan:1998we}. It does not occur at LO, because the total neutron-proton spin $S=1$ is conserved if all interactions are of $s$-wave type. In the presence of the $p$-wave state $\Iso{Be}{11}^\ast$, however, $S$ may change, and transitions \mbox{$\ket{\d}\rightarrow\ket{\pi}\rightarrow\ket{\Iso{Be}{10}+\text{np}(\spec{1}{S}_0)}$} become possible; see Fig.~\ref{Fig:VirtualState}. However, the virtual state is not only suppressed due to the intermediate $\ket{\pi}$ channel. Since multiple spin changes [\mbox{$\sim(\Rc/\Rh)^2$} or smaller] are negligible at NLO, a virtual state leads to $S=0$ in the final state of \reaction . The corresponding phase space is $1/3$ the size of $S=1$, yielding a suppression of \mbox{$\Rc/(3\Rh)\lesssim (\Rc/\Rh)^2$} (N${}^2$LO).

Neutron-proton $p$-wave interactions are of order N${}^3$LO. The reason is the lack of a shallow neutron-proton $p$-wave bound or resonance state. In the $\Iso{Be}{11}^\ast$ sector, $iG_\pi$ approaches the large scattering volume $a_\pi=\unit[(457\pm67)]{fm^3}$ for small $\Ecm$ \cite{Typel:2004zm}. This large value is a consequence of the small binding momentum $\gamma_\pi$ since $a_\pi\sim 2r_\pi^{-1}\gamma_\pi^{-2}\sim \Rc\Rh^2$; see Eq.~\eqref{Eq:PoleExp_pi}. Scattering volumes in the neutron-proton channels $\spec{3}{P}_2$, $\spec{3}{P}_1$, $\spec{3}{P}_0$, and $\spec{1}{P}_1$ are much smaller. Using the Nijmegen partial wave analysis for N-N scattering of Ref.~\cite{Stoks:1994wp}, we have checked that they are all of the natural size \mbox{$r_\text{d}^3\approx\unit[5.36]{fm^3}\sim \Rc^3\ll a_\pi$} or smaller (N${}^3$LO). In fact, the $p$-wave phase shifts themselves are suppressed compared to the $\spec{3}{S}_1$ phase shift. Even for the maximal neutron-proton center-of-mass energy $\Ecm=\unit[15.61]{MeV}$ available in the experiment by Schmitt \etal, the suppression is of the order $0.09\sim(\Rc/\Rh)^3$ (N${}^3$LO).

In Ref.~\cite{Goosman:1970vn}, several boron-11 resonances have been observed in \mbox{$\Iso{Be}{10}(\text{p},\,\gamma)\Iso{B}{11}$}; see Fig.~\ref{Fig:Levels}. The lowest one ($1/2^+$) occurs at a proton-core center-of-mass energy \mbox{$E_\text{r}=\unit[(1.33\pm 0.04)]{MeV}$}. It has a total width \mbox{$\Gamma=\unit[(230\pm 65)]{keV}$} and the branching ratio for decay into \mbox{$\Iso{Be}{10}+\text{p}$} is close to $1$ \cite{Goosman:1970vn}. The resonance represents a pole at \mbox{$\Ecm=E_\text{r}-i\Gamma/2$} in the Coulomb-modified
resonance propagator; see for example Refs.~\cite{Kok:1982dr, Kong:1998sx}. This pole position implies effective range terms \mbox{$a_\text{C}^{-1}=(\unit[(-2.7\pm 0.8)]{fm})^{-1}$} and \mbox{$r_\text{C}/2\,(2\muNc E_\text{r})= (\unit[(-3.5\pm1.4)]{fm})^{-1}$}, which scale like $\Rc^{-1}$. Moreover, in three-body diagrams, the resonance propagator comes along
with a Gamow-Sommerfeld factor \mbox{$0<C_{\eta}^2<1$} \cite{Kong:1999mp}. It gives the probability of two charged particles to meet in one point. At resonance, it takes the small value \mbox{$0.13\lesssim(\Rc/\Rh)^2$}. It follows that the influence of the resonance propagator
on the reaction is suppressed by three orders in $\Rc/\Rh$ compared to $iG_\sigma$ (N${}^3$LO).
Note that there are more boron-11 states around $E=0$, which could possibly couple strongly to the proton-core system. However, transitions to those states would involve even smaller Gamow factors \mbox{$C_{\eta}^2<(\Rc/\Rh)^2$}. Thus, we neglect strong proton-core interactions at NLO.
 
During the reaction process, the $\Iso{Be}{11}$ state could break up into an excited core $\Iso{Be}{10}^\ast$ and a neutron. Thus, $\ket{\sigma}$ in principle couples to the additional intermediate channel \mbox{$\ket{\Iso{Be}{10}^\ast+\text{d}}$} via neutron exchanges. However, each such channel comes along with two couplings of order $\Rc^2$; see Appendix~\ref{App:CoreExc} for details. Thus, dynamical core excitations can be neglected at NLO.

Diagrams involving direct photon couplings to the auxiliary fields $d_i$ and $\sigma_\alpha$ do not enter before N${}^2$LO. They are one order smaller than the bubble diagrams \cite{Konig:2015aka}, which are de facto of NLO; see above. The Coulomb field could also induce E$1$ transitions between $\Iso{Be}{11}$ and $\Iso{Be}{11}^\ast$ (or between the deuteron and a neutron-proton $p$-wave channel) \cite{Hammer:2011ye}. Such a contribution is shown in Fig.~\ref{Fig:VabExc} (d). It is negligible at NLO due to the subleading nature of the $\Iso{Be}{11}^\ast$ propagator and due to the photon propagator, which is governed by the large external momentum scale $p$.

\section{Summary and outlook}
\label{Sec:SummaryOutlook}

In this work, we carried out the first Halo EFT calculation of
deuteron-induced transfer reactions. As a working example, we considered \reaction, involving the one-neutron halo nucleus $\Iso{Be}{11}$. The degrees of freedom in this approach are the $\Iso{Be}{10}$ core, the neutron, and the proton. Strong interactions are described by contact forces alone. To obtain the differential cross section, the reaction amplitude was constructed diagrammatically in an expansion in the ratio \mbox{$\Rc/\Rh\sim 0.4$} of core and halo radius. The corresponding Faddeev equation contains all dynamical features of a transfer reaction including two-body breakup contributions. A three-body force ensures internal consistency. We included the Coulomb force by considering the dominant photon exchange diagrams, which were iterated to all orders in the Faddeev equation.

The differential cross section was compared to experimental data by Schmitt \etal\ \cite{Schmitt:2012bt, Schmitt:2013uye}. In agreement with Yang and Capel \cite{Yang:2018nzr}, who calculated the cross section in the adiabatic distorted wave approximation, we found that Halo EFT is able to describe scattering at low beam energies \mbox{$\Ed\lesssim \unit[15]{MeV}$} (center-of-mass energies \mbox{$E\lesssim\unit[10]{MeV}$}). In this regime, the reaction can be considered peripheral, i.e., it predominantly depends on the long-range tail of the $\Iso{Be}{11}$ wave function. This part is systematically reproduced by the $\Rc/\Rh$ expansion.

Our theory contains only few information on the spectra of the involved particles. We included, in particular, only two-body states with a binding momentum $\gamma$ clearly smaller than the respective momentum scale of short-range physics; see Fig.~\ref{Fig:Levels}. The influence of such states should be enhanced by powers of $\gamma^{-1}$ compared to those far away from the two-body threshold. As a consequence of this reduction, we were able to describe data using only a minimal amount of experimental input. At LO in the $\Rc/\Rh$ expansion, only the binding energies of deuteron and $\Iso{Be}{11}$ are needed; see Table~\ref{Tab:InputParameters}. NLO corrections arise from respective effective ranges and the first exited state $\Iso{Be}{11}^\ast$. The effective ranges of the $\Iso{Be}{11}$ states were extracted from the ANCs of the \textit{ab~initio} calculation by Calci \etal\ \cite{Calci:2016dfb}. Both NLO corrections modify the cross section at a $\unit[40]{\%}$ level, as predicted by the power counting.

While our results describe data at \mbox{$E_\text{d}\lesssim \unit[15]{MeV}$} fairly well, they strongly overestimate the cross section at higher beam energies. Apparently, the low-energy expansion of Halo EFT converges, if at all, slowly at these energies. In order to improve the expansion, it might be necessary to modify the three-body power counting, which, at the moment, counts loop momenta like small binding momenta. In a more sophisticated power counting, tailored to beam energies $\Ed>\unit[12]{MeV}$, neglected higher-order interactions might already occur at lower orders. Such a scheme should be developed in the future. Hints on missing ingredients can be inferred from previous theoretical analyses, e.g., by Schmitt \etal\ in Ref.~\cite{Schmitt:2013uye}, Deltuva \etal\ in Ref.~\cite{Deltuva:2016his}, or Yang and Capel in Ref.~\cite{Yang:2018nzr}, which were successful in describing also scattering for \mbox{$E_\text{d}\geq \unit[15]{MeV}$}. The model used in Ref.~\cite{Yang:2018nzr} contains the same amount of information on the $\Iso{Be}{11}$ spectrum as our work. Thus, we do not expect the inclusion of beryllium-11 levels beyond the first excited state to be of prime importance.

Instead, core excitations following $\Iso{Be}{11}$ breakup and two-body interactions in higher partial waves might provide enough absorption to lower cross sections at higher energies. Moreover, we might need to consider not explicitly measured loss channels, in particular due to deep boron-11 states indicated in Fig.~\ref{Fig:Levels}, at these energies. Usually, such effects are included using optical model potentials, adjusted to, for example, proton-core scattering data. In the future, we will instead introduce imaginary contact terms to the strong Lagrangian, a method called ``Open EFT'' \cite{Braaten:2016sja}. It was applied successfully to a broad range of inelastic processes including quarkonium decays in nonrelativistic QCD \cite{Bodwin:1994jh} and three-body recombinations of ultracold atoms \cite{Braaten:2001hf}.

Let us emphasize again, that Halo EFT is ideally suited for the description of strong interactions at low energies. In this sense, our long-term goal is to apply the developed framework to the astrophysical regime. While Coulomb effects become nonperturbative then, short-range effects should become less important. In this context, it will be interesting to calculate the cross section for \reaction$^\ast$, which was measured in by Schmitt \etal\ \cite{Schmitt:2012bt, Schmitt:2013uye}. This process is less peripheral than \reaction\ \cite{Yang:2018nzr}, which is why naive application of the current power counting at experimental energies leads to an overestimation of the data. At very small energies, however, the power counting should be appropriate. Note, however, that certain Coulomb diagrams involving $\Iso{Be}{11}^\ast$, which we could neglect for \reaction, would become important for \reaction$^\ast$.
Moreover, we could apply the framework to other deuteron-induced reactions like \mbox{$\Iso{C}{14}(\text{d},\,\text{p})\Iso{C}{15}$}.

\begin{acknowledgments}
	We thank D. R. Phillips for giving valuable
	feedback on the manuscript and S. K\"onig for providing
	information on the calculation of the Coulomb box
	diagrams. M. S. appreciates stimulating discussions with
	I. Thompson, D. Baye, and other participants of the INT
	Program INT-17-1a
	``Toward Predictive Theories of Nuclear Reactions Across the
	Isotopic Chart.'' Moreover, M. S. sincerely thanks the
	Nuclear Theory groups of
	UT Knoxville and Oak Ridge National Laboratory for their kind
	hospitality and support during his research stay.
	This work has been funded by the Deut\-sche
	For\-schungs\-ge\-mein\-schaft (DFG, German Research
	Foundation), Pro\-jekt\-num\-mer 279384907, SFB 1245,
	by the National Science Foundation under Grant No.
	PHY-1555030, by the Bundesministerium f\"ur Bildung und Forschung (BMBF) through Contract No. 05P18RDFN1, and by the Office of Nuclear Physics, U.S.
	Department of Energy under Contract No. DE-AC05-00OR22725.
\end{acknowledgments}

\appendix

\section{Core excitation effects}
\label{App:CoreExc}

In this section, we show that core excitation effects in the pole region are taken care of in this work due to renormalization onto low-energy observables. For that, we consider a theory with an explicit $\Iso{Be}{10}^\ast$ field $C_m$ \mbox{($m\in\{-2,\,\dots\,,2\}$)} by adding a piece
\begin{equation}
	\label{Eq:L1Be10Exc}
	\mathcal{L}_{1,\Iso{Be}{10}^\ast}
	=C_m^\dagger\left[
		\partial_0 + \frac{\bs{\nabla}^2}{2\mc}-\Ex
	\right]C_m
\end{equation}
to the Lagrangian. A similar approach has been chosen by Zhang \etal\ to analyze effects of the core excitation $\Iso{Li}{7}^\ast$  on the \mbox{$\Iso{Li}{7}(\text{n},\,\gamma)\Iso{Li}{8}$} reaction \cite{Zhang:2013kja}. Moreover, Zhang \etal\ and Ryberg \etal\ used a $\Iso{Be}{7}^\ast$ core excitation field in their calculations of the $S$-factor of \mbox{$\Iso{Be}{7}(\text{p},\,\gamma)\Iso{B}{8}$} \cite{Zhang:2014zsa, Ryberg:2014exa}. In both systems, the core excitation occurs at low energies. That, however, is not true in our case where \mbox{$(2\muNc\Ex)^{1/2}\sim \Rc^{-1}$} is large.

Together with a neutron, $\Iso{Be}{10}^\ast$ couples to the $\Iso{Be}{11}$ ground state in a $d$-wave. In terms of the redefined field $\tilde{\sigma}_{\alpha}$, we thus write
\begin{equation}
	\label{Eq:L2Be10Exc}
	\mathcal{L}_{2,\Iso{Be}{10}^\ast}
	=-\!\!\!\!\sum_{s\in\{3/2,\,5/2\}}\ \frac{g_{\sigma,\text{x}}^{(s)}}{g_\sigma}\,
	\Cl{1/2\alpha}{2m}{sm_s}\Cl{2m_l}{sm_s}{1/2 \alpha'}
	\left[
		\tilde{\sigma}_{\alpha'}^\dagger
		\left(
			n_\alpha \left\lbrace
				-i\overleftrightarrow{\bs{\nabla}}
			\right\rbrace_{2m_l}C_m
		\right)+\text{H.c.}
	\right].
\end{equation}
The vertex term contains a Galilei-invariant derivative \mbox{$\overleftrightarrow{\bs{\nabla}}\equiv
\muNc(\mN^{-1}\,\overleftarrow{\bs{\nabla}}-\mc^{-1}\,\overrightarrow{\bs{\nabla}})$}. It is embedded in the tensor structure
\begin{equation}
	\label{Eq:LWaveTensor}
	\left\lbrace
		\bs{\mathcal{O}}
	\right\rbrace_{lm_l}
	\equiv \sqrt{\frac{4\pi}{2l+1}}\,\left|
		\bs{\mathcal{O}}
	\right|^l\,
	Y_{l}^{m_l}\left(\hat{\bs{\mathcal{O}}}\right)
\end{equation}
with $l=2$, where $Y_l^{m_l}(\hat{\bs{\mathcal{O}}})$ denotes a spherical harmonic, evaluated at \mbox{$\hat{\bs{\mathcal{O}}}\equiv \bs{\mathcal{O}}/|\bs{\mathcal{O}}|$}.

The mass difference \mbox{$\Ex+B_\sigma\gg B_\sigma$} in the transition is of natural size. Thus, we assume no fine-tuning in this scattering channel and count \mbox{$g_{\sigma,\text{x}}^{(s)}\sim \Rc^{3/2}$}. It follows that the overall couplings \mbox{$g_{\sigma,\text{x}}^{(s)}/g_\sigma\sim \Rc^2$} are natural as well, since \mbox{$g_\sigma\sim r_\sigma^{-1/2}\sim \Rc^{-1/2}$}; see Eq.~\eqref{Eq:Matching_sigma2}.

The core excitation modifies the $\Iso{Be}{11}$ propagator through the $\Iso{Be}{10}^\ast$-neutron self-energy loop \mbox{$-i\Sigma_{\sigma,\text{x}}\,\delta^{\alpha\alpha'}$}. It resembles the $\Iso{Be}{10}$-neutron self-energy loop of Fig.~\ref{Fig:Dyson_sigma}, but the core line has to be replaced by a core excitation line. Using the PDS scheme, we find
\begin{align}
	\nonumber
	\Sigma_{\sigma,\text{x}}(\Ecm)
	=&\ -\sum_s\left(\frac{g_{\sigma,\text{x}}^{(s)}}{g_\sigma}\right)^2\frac{\muNc}{10\pi}\,
	\left[
		2\muNc(\Ecm-\Ex+i\epsilon)
	\right]^2
	\\
	&\hspace{2cm}\times\left(
		\Lambda_\text{PDS}-\left[
			-2\muNc(\Ecm-\Ex+i\epsilon)
			\right]^{1/2}
	\right)
	\\
	\equiv&\ -g_\sigma^{-2}\sum_n \Delta_{\sigma,\text{x}}^{(n)}(\Ecm+i\epsilon)^n.
\end{align}
Note that $\Sigma_{\sigma,\text{x}}$ is analytic for \mbox{$\Ecm<\Ex$}, i.e., it can be expanded at $\Ecm=0$. The resulting coefficients $\Delta_{\sigma,\text{x}}^{(n)}$ then contribute to the unrenormalized parameters $\Delta_\sigma^{(n)}$ \mbox{($\Delta_\sigma^{(1)}\equiv -1$)} of the bare $\Iso{Be}{11}$ propagator; see Eq.~\eqref{Eq:Lagrangian-2b}. Thus, renormalization onto observables $\gamma_\sigma$ (or $a_\sigma$), $r_\sigma$, etc. automatically takes care of core excitation effects at small $\Ecm$, where the pole is located. In other words, $C_m$ does not introduce any new information to the two-body sector and can be integrated out.

\section{Partial wave expansion}
\label{App:PWProj}

Let us consider a general interaction $\mathcal{I}$, which could be an amplitude $T$, a neutron exchange potential $V$ or a Coulomb diagram interaction $\Gamma$. We expand $\mathcal{I}$ in
tensor spherical harmonics
\begin{equation}
	\label{Eq:VecSphHarm}
	\left(\bs{Y}_{(L,S)Jm_J}(\hat{\bs{p}})\right)^m
	\equiv \sum_{m_L}
	\Cl{Lm_L}{Sm}{Jm_J}
	Y_{L}^{m_L}(\hat{\bs{p}})
\end{equation}
by writing
\begin{align}
	\label{Eq:PartialWaveExpansion}
  	\mathcal{I}^{Sm,S'm'}\left(\bs{p},\bs{q};\,E\right)
  	=&\ \sum_J \sum_{L,L'}
    \mathcal{I}^{\spec{2S+1}{L}_J,\spec{2S'+1}{L'}_J}\left(p,\,q;\,E\right)
    P_{\,\spec{2S+1}{L}_J,\spec{2S'+1}{L'}_J}^{\,m,m'}\left(\hat{\bs{p}},\,\hat{\bs{q}}\right),
  \\
  P_{\,\spec{2S+1}{L}_J,\spec{2S'+1}{L'}_J}^{\,m,m'}\!\left(\hat{\bs{p}},\,\hat{\bs{q}}\right)
   \equiv&\ 4\pi\sum_{m_J}
     \left(
     \bs{Y}_{(L,S)Jm_J}\left(\hat{\bs{p}}\right)
     \right)^m
     \left(\bs{Y}_{(L',S')Jm_J}
     \left(\hat{\bs{q}}\right)\right)^{m'\,\ast}\,.
\end{align}
Specific partial waves can be extracted via
\begin{align}
  \mathcal{I}^{\spec{2S+1}{L}_J,\spec{2S'+1}{L'}_J}\left(p,\,q;\,E\right)
  =
  \frac{(4\pi)^{-2}}{2J+1}\sum_{m,m'}\int_{\Omega_{\bs{p}},\Omega_{\bs{q}}}
  P_{\,\spec{2S'+1}{L'}_J,\spec{2S+1}{L}_J}^{\,m',m}\!\left(\hat{\bs{q}},\,\hat{\bs{p}}\right)
  \,
  \mathcal{I}^{Sm,S'm'}\left(\bs{p},\bs{q};\,E\right).
\end{align}

\section{Coulomb diagrams}
\label{App:CoulDiags}
The Coulomb diagrams in Fig.~\ref{Fig:CoulombDiagrams} resemble such considered by K\"onig \etal\ for the three-nucleon system \cite{Konig:2015aka}. However, they exhibit nontrivial dependencies on the mass ratio \mbox{$y\equiv\mN/\mc$}. The bubble interactions read
\begin{align}
  \nonumber
  \Gamma_{\d\d}^{1m,1m'}
  \left(\bs{p},\,\bs{q};\,E\right)
  =&\ \delta^{mm'}\,\frac{
     Q_\text{c}\alpha\,\mN^2
     }{
     \left(\bs{p}-\bs{q}\right)^2
     +\lambda^2-i\epsilon
     }
  \\
  \label{Eq:Gamma_dd}
   &\ \times
   f\Big(
       \bs{p}-\bs{q},\,\mathcal{A}_\d(p;\,E),\,\mathcal{A}_\d(q;\,E)
   \Big)\,,
  \\
  \nonumber
  \Gamma_{\sigma\sigma}^{Sm,S'm'}
  \left(\bs{p},\,\bs{q};\,E\right)
  =&\ \delta^{SS'}\delta^{mm'}\,\frac{
     Q_\text{c}\alpha\,(2\muNc)^2
     }{
     	\left(\bs{p}-\bs{q}\right)^2
     	+\lambda^2-i\epsilon
     }
  \\
  \label{Eq:Gamma_ss}
   &\hspace{-0.45cm}\ \underbrace{
     \times
     f\Bigg(
         \frac{y}{\xi}\left(\bs{p}-\bs{q}\right),\,\mathcal{A}_\sigma(p;\,E),\,\mathcal{A}_\sigma(q;\,E)
         \Bigg)
         \,,
         }_{=\big(
         \sqrt{\mathcal{A}_\sigma(q;\,E)}-\sqrt{\mathcal{A}_\sigma(p;\,E)}
         \big)/\big(
         \mathcal{A}_\sigma(q;\,E)-\mathcal{A}_\sigma(p;\,E)
         \big)
         +\OO{y^2}
         }
\end{align}
and the box interactions are given by
\begin{align}
  \nonumber
  \Gamma_{\sigma\d}^{Sm,1m'}
  \left(\bs{p},\,\bs{q};\,E\right)
  =&\ -Q_\text{c}\alpha\,\mN\,
     V_{\sigma\d}^{Sm,1m'}\left(\bs{p},\,\bs{q};\,E\right)
  \\
  \nonumber
   &\ \times\Bigg[
     f\Bigg(
     \bs{p}-y\bs{q},\,\xi^2 \mathcal{A}_\sigma(p;\,E),\,\mathcal{A}_\d(q;\,E)
     \Bigg)
	\\
  \label{Eq:Gamma_sd}
   &\hspace{4em}
     -\frac{\lambda}{
     \bs{p}\cdot\bs{q} + p^2+\xi{q}^2
     -\mN(E+i\epsilon)
     }
     +\OO{\lambda^2}
     \Bigg]\,,
  \\
  \label{Eq:Gamma_ds}
  \Gamma_{\d\sigma}^{1m,S'm'}
  \left(\bs{p},\,\bs{q};\,E\right)
  =&\ \Gamma_{\sigma\d}^{S'm',1m}
     \left(\bs{q},\,\bs{p};\,E\right),
\end{align}
where we defined \mbox{$\xi\equiv(1+y)/2$}. Moreover, \mbox{$\alpha\equiv e^2/(4\pi)\approx 1/137$} is the fine structure constant and \mbox{$Q_\text{c}=4$} is the core charge. All interactions involve the function
\begin{equation}
	f\left(\bs{\Delta},\,\mathcal{A}_1,\,\mathcal{A}_2\right)
	\equiv
	\frac{1}{
		\left|\bs{\Delta}\right|
	}\,
	\text{tan}^{-1}\!\left(
		\frac{
			\mathcal{A}_1-\mathcal{A}_2
			+\bs{\Delta}^2/4
		}{
			\left|\bs{\Delta}\right|
			\sqrt{\mathcal{A}_2}
		}
	\right)
	+\left[\mathcal{A}_1\leftrightarrow \mathcal{A}_2\right],
\end{equation}
whose arguments involve the expressions
\begin{align}
	\mathcal{A}_\d(p;\,E)\equiv&\ \frac{1+2y}{4}\,p^2-\mN (E+i\epsilon)
	\xrightarrow{\text{\ on shell\ }}\gamma_\d^2\,,
	\\
	\mathcal{A}_\sigma(p;\,E)\equiv&\ \xi^{-2}\,\frac{1+2y}{4}\,p^2-\xi^{-1}\mN (E+i\epsilon)
	\xrightarrow{\text{\ on shell\ }}\gamma_\sigma^2\,.
\end{align}

The form of $\Gamma_{\sigma\sigma}$ can be simplified
significantly by neglecting terms of order $\OO{y^2}$; see
Eq.~\eqref{Eq:Gamma_ss}. This approximation is justified since
\mbox{$y^2=0.01$} is a tiny number. The only angular dependence then comes
from the photon propagator, which can be projected onto certain
partial waves analytically.

The bubble diagrams $-i\Gamma_{\d\d}^{1m,1m'}$ and
$-i\Gamma_{\sigma\sigma}^{Sm,S'm'}$ are linear in the Coulomb
propagator. Thus, their largest contributions to the transfer reaction
comes from the region of small momentum transfers $\bs{p}-\bs{q}$. For
$\bs{p}=\bs{q}$, the values of the function
$f$ in Eqs.~\eqref{Eq:Gamma_dd}--\eqref{Eq:Gamma_ss} collapse to
\mbox{$[\mathcal{A}_a(p;\,E)]^{-1/2}/2\xrightarrow{\ \text{on shell}\ }1/(2\gamma_a)$}
\mbox{($a\in\{\d,\,\sigma\}$)}. Thus, the deuteron and halo loops of the LO
bubble diagrams in Fig.~\ref{Fig:CoulombDiagrams} may be counted like $\mN/\gamma$.

The Coulomb diagram interactions $\Gamma_{ab}$ can be connected to the
$s$-wave projected functions $K_\text{bubble}$ and $K_\text{box}$ of
Ref.~\cite{Konig:2015aka} by taking the limits
\mbox{$y,Q_\text{c}\rightarrow 1$}. We find
\begin{align}
	\left.
		\int_{-1}^1\! \d x\ 
		\Gamma_{aa}^{10,10}
		\left(\bs{p},\,\bs{q};\,E\right)
	\right|_{y,Q_\text{c}\rightarrow 1}
	=&\ -\frac{\mN}{4\pi}\,K_\text{bubble}\left(E;\,p,\,q\right)\ (a\in\{\d,\,\sigma\}),
	\\
	\left.
		\int_{-1}^1\! \d x\ 
		\Gamma_{\sigma\d}^{10,10}
		\left(\bs{p},\,\bs{q};\,E\right)
	\right|_{y,Q_\text{c}\rightarrow 1}
	=&\ -\frac{\mN}{2\pi}\,K_\text{box}\left(E;\,p,\,q\right),
\end{align}
where \mbox{$x\equiv \bs{p}\cdot\bs{q}/(pq)$}.

\section{Excited state of beryllium-11}
\label{App:Be11Exc}

In this section, we discuss the inclusion of the excited state $\Iso{Be}{11}^\ast$ at NLO in the reaction calculation. The Lagrangian part
\begin{align}
	\nonumber
	\mathcal{L}_{\Iso{Be}{11}^\ast}
	=&\ \pi_\alpha^\dagger \left[
		\Delta_\pi^{(0)}
		+\left(
			i\partial_0+\frac{\bs{\nabla}^2}{2M_\text{Nc}}
		\right)
	\right]\pi_\alpha
	\\
	\label{Eq:LBe11Exc}
	&\ -g_\pi\ \Cl{1/2\,\alpha}{1m_l}{1/2\,\alpha'}
	\left[
		\pi_{\alpha'}^\dagger \left(
			n_\alpha \left\lbrace
				-i\overleftrightarrow{\bs{\nabla}}
			\right\rbrace_{1m_l}\, c
		\right) + \text{H.c.}
	\right]+\cdots
\end{align}
of Eq.~\eqref{Eq:Lagrangian-2b} contains an auxiliary field
$\pi_\alpha$ \mbox{($\alpha\in\{-1/2,\,1/2\}$)} for $\Iso{Be}{11}^\ast$ with
renormalization-dependent parameters $\Delta_\pi^{(0)},\,g_\pi\in\mathbb{R}$. The Galilei-invariant derivative $\overleftrightarrow{\bs{\nabla}}$ and the $p$-wave tensor structure $\{\bs{\mathcal{O}}\}_{1m_l}$ are defined in Appendix~\ref{App:CoreExc}. Unlike
in the $s$-wave case, both the constant and derivative part of the
bare propagator term in Eq.~\eqref{Eq:LBe11Exc} are needed to describe
the shallow $p$-wave state \cite{Bertulani:2002sz, Hammer:2011ye}. The full $\Iso{Be}{11}^\ast$ propagator can be obtained by resumming
all two-body loops, similarly to Fig.~\ref{Fig:Dyson_sigma}. For more details, we refer to Ref.~\cite{Hammer:2011ye}. After
proper renormalization and field redefinitions
\mbox{$\pi_\alpha^{(\dagger)}\rightarrow
\tilde{\pi}_\alpha^{(\dagger)}\equiv g_\pi\,\pi_\alpha^{(\dagger)}$},
the propagator $G_\pi$ around the pole at $\Ecm=-B_\pi$ is given by Eq.~\eqref{Eq:PoleExp_pi}.

In the NLO three-body system, the intermediate state \mbox{$\ket{\pi}\equiv\ket{\text{p}+\Iso{Be}{11}^\ast}$} couples to $\ket{\d}$ via neutron exchange potentials shown in Fig.~\ref{Fig:VabExc}. They read
\begin{align}
	\nonumber
	V_{\pi\d}^{Sm,1m'}\left(\bs{p},\,\bs{q};\,E\right)
	=&\ \mN\,\sqrt{6}\left\lbrace
		\begin{matrix}
			S&1&1\\
			1/2&1/2&1/2
		\end{matrix}
	\right\rbrace
	\\
	&\times \frac{
		\sum_{m_l}\Cl{1m_l}{1m'}{Sm}\left\lbrace\frac{1}{1+y}\,\bs{p}+\bs{q}\right\rbrace_{1m_l}^\ast
	}{
		\bs{p}\cdot\bs{q} + p^2+\frac{1+y}{2}{q}^2-\mN(E+i\epsilon)
	}\,,
	\\
	V_{\d\pi}^{1m,S'm'}\left(\bs{p},\,\bs{q};\,E\right)
	=\ &\left[
		V_{\pi d}^{S'm',1m}\left(\bs{q},\,\bs{p};\,E\right)
	\right]^\ast
\end{align}
with \mbox{$S\in\{0,\,1\}$} in the $\ket{\pi}$ channel
and involve a $6j$ symbol. Partial wave projections are given by
\begin{align}
	\nonumber
	V_{\pi\d}^{\spec{2S+1}{L}_J\,\spec{3}{L'}_J}\left(p,\,q;\,E\right)
	=&\ (-1)^{J+1}\sqrt{2\,(2S+1)(2L+1)(2L'+1)}
	\\
	\nonumber
	&\hspace{-1cm}\times\Cl{L0}{L'0}{10}\,\left\lbrace
		\begin{matrix}
			S&1&1\\
			1/2&1/2&1/2
		\end{matrix}
	\right\rbrace
	\left\lbrace
		\begin{matrix}
			S&1&1\\
			L'&L&J
		\end{matrix}
	\right\rbrace
	\\
	\label{Eq:Vpid}
	&\hspace{-1cm}\times\frac{\mN}{pq}\left[
		\frac{1}{1+y}\,p\,Q_{L'}+q\,Q_L
	\right]\left(
		-\frac{p^2+\frac{1+y}{2}q^2-\mN(E+i\epsilon)}{pq}
	\right),
	\\
	\label{Eq:Vdpi}
	V_{\d\pi}^{\spec{3}{L}_J\,\spec{2S'+1}{L'}_J}\left(p,\,q;\,E\right)
	=&\ V_{\pi\d}^{\spec{2S'+1}{L'}_J\,\spec{3}{L}_J}\left(q,\,p;\,E\right).
\end{align}
A direct transition potential between $\ket{\sigma}$ and $\ket{\pi}$
is not induced by the Lagrangian, i.e., these states can only be
connected via an intermediate state $\ket{\d}$.

The Clebsch-Gordan coefficient and the $6j$ symbols in
Eq.~\eqref{Eq:Vpid} imply some selection rules. First, only
transitions with \mbox{$|\Delta L|=1$} are allowed. It follows that for
$J=0$, we have \mbox{$L_\d=L_\sigma=1, L_\pi=0$}, and for fixed \mbox{$J\geq 1$}, the
system decouples into the two subsystems (1)
\mbox{$L_\d=L_\sigma=J,\,L_\pi=J\pm 1$} and (2)
\mbox{$L_\d=L_\sigma=J\pm 1,\,L_\pi=J$}.  Second, \mbox{$S_\pi=1$} is fixed in
subsystem (1), while both options \mbox{$S_\pi\in\{0,\,1\}$} are allowed in
subsystem (2). Last, in subsystem (2), the two channels
\mbox{$L_\d=L_\sigma=J\pm 1$} further decouple after defining rotated spin
states
\begin{equation}
	\label{Eq:RotatedPiStates}
	\begin{pmatrix}
		\ket{\pi,\,\spec{\bar{3}}{J}_J}\\[1ex]
		\ket{\pi,\,\spec{\bar{1}}{J}_J}
	\end{pmatrix}
	\equiv
	\frac{1}{\sqrt{2J+1}}
	\begin{pmatrix}
		\sqrt{J+1} & \sqrt{J} \\ -\sqrt{J} & \sqrt{J+1}
	\end{pmatrix}
	\begin{pmatrix}
		\ket{\pi,\,\spec{3}{J}_J}\\[1ex]
		\ket{\pi,\,\spec{1}{J}_J}
	\end{pmatrix}\,.
\end{equation}
Note that \mbox{$\bar{3}= 3$} and \mbox{$\bar{1}= 1$} for $J=0$. The corresponding partial wave potentials read
\begin{align}
	\nonumber
	V_{\pi\d}^{\spec{\overline{2\mp1}}{J}_J,\spec{3}{\,(J\pm1)}_J}(p,\,q;\,E) =&\ 	\mp\frac{1}{\sqrt{3}}
	\\
	&\hspace{-2cm}\times\frac{\mN}{pq}\left[
		\frac{1}{1+y} p\, Q_{J\pm1} + q\, Q_J
	\right]\left(
		-\frac{p^2+\frac{1+y}{2}q^2-\mN(E+i\epsilon)}{pq}
	\right)\,,
	\\
	V_{\d\pi}^{\spec{3}{\,(J\pm1)}_J,\spec{\overline{2\mp1}}{J}_J}(p,\,q;\,E)
	=&\ V_{\pi\d}^{\spec{\overline{2\mp1}}{J}_J,\spec{3}{\,(J\pm1)}_J}(q,\,p;\,E)\,.
\end{align}

\begin{table}
	\begin{tabular}{p{0.15\textwidth}p{0.15\textwidth}p{0.15\textwidth}p{0.15\textwidth}p{0.3\textwidth}}\hline\hline
		Subsystem & $L_\d=L_\sigma$ & $S_\d=S_\sigma$ & $L_\pi$ & $S_\pi$\\\hline
		(1) & $J$ & $1$ & $J\pm1$ & $1$\\
		(2a) & $J-1$ & $1$ & $J$ & $\bar{3}\propto \sqrt{J+1}\times 3+\sqrt{J}\times 1$\\
		(2b) & $J+1$ & $1$ & $J$ & $\bar{1}\propto -\sqrt{J}\times 3+\sqrt{J+1}\times 1$\\\hline\hline
	\end{tabular}
	\caption{\label{Tab:NLOSubsystems}Subsystems of fixed $J$
          after including the excited state channel
          $\ket{\pi}$. Subsystems (1) and (2a) require $J\geq 1$. The
          quantum numbers $\bar{3}$ and $\bar{1}$ in subsystems (2a)
          and (2b) refer to rotated spin states of $\ket{\pi}$; see
          Eq.~\eqref{Eq:RotatedPiStates}.}
\end{table}

In summary, for fixed $J\geq 1$, we find the three decoupled
subsystems (1), (2a), and (2b) presented in
Table~\ref{Tab:NLOSubsystems}. Just as in the LO case, they can be
identified by the conserved quantum number (1) \mbox{$L_\d=J$}, (2a)
\mbox{$L_\d=J-1$}, and (2b) \mbox{$L_\d=J+1$}. In the case \mbox{$J=0$}, only system (2b) is
allowed.

\section{NLO equations}
\label{App:NLOEquations}

As explained in Appendix~\ref{App:Be11Exc}, the introduction of the
excited state $\Iso{Be}{11}^\ast$ produces three decoupled scattering
systems for fixed $J\geq 1$, corresponding to
\mbox{$L_\d=L_\sigma\in\{J-1,\,J,\,J+1\}$}, and a single system for $J=0$
with \mbox{$L_\d=L_\sigma=1$}. The respective NLO amplitude vectors
$\vec{T}^{\text{(NLO)}\,[L_\d,J]}$ read
\begin{align}
	\label{Eq:TNLOVec1}
	\vec{T}^{\text{(NLO)}\,[J,J]}
	=&\ \begin{pmatrix}
		T_{\d\d}^{\text{(NLO)}\,\spec{3}{J}_J,\spec{3}{J}_J}\\[0.5ex]
		T_{\sigma\d}^{\text{(NLO)}\,\spec{3}{J}_J,\spec{3}{J}_J}\\[0.5ex]
		T_{\pi\d}^{\spec{3}{(J-1)}_J,\spec{3}{J}_J}\\[0.5ex]
		T_{\pi\d}^{\spec{3}{(J+1)}_J,\spec{3}{J}_J}
	\end{pmatrix}\ (J\geq 1)\,,
	\\
	\label{Eq:TNLOVec2}
	\vec{T}^{\text{(NLO)}\,[J\pm 1,J]}
	=&\ \begin{pmatrix}
		T_{\d\d}^{\text{(NLO)}\,\spec{3}{(J\pm1)}_J,\spec{3}{(J\pm1)}_J}\\[0.5ex]
		T_{\sigma\d}^{\text{(NLO)}\,\spec{3}{(J\pm1)}_J,\spec{3}{(J\pm1)}_J}\\[0.5ex]
		T_{\pi\d}^{\spec{\overline{2\mp1}}{J}_J,\spec{3}{(J\pm1)}_J}
	\end{pmatrix}\ (J\geq 0\ \text{and}\ J\geq 1)\,.
\end{align}
They are determined by the interaction and propagator matrices
\begin{align}
	\underline{\underline{K}}^{\text{(NLO)}\,[J,J]}
	=&\ \begin{pmatrix}
		\Gamma_{\d\d}^{\spec{3}{J}_J,\spec{3}{J}_J} & (V_{\d\sigma}+\Gamma_{\d\sigma})^{\spec{3}{J}_J,\spec{3}{J}_J} & V_{\d\pi}^{\spec{3}{J}_J,\spec{3}{(J-1)}_J} & V_{\d\pi}^{\spec{3}{J}_J,\spec{3}{(J+1)}_J}
		\\[0.5ex]
		(V_{\sigma\d}+\Gamma_{\sigma\d})^{\spec{3}{J}_J,\spec{3}{J}_J} & \Gamma_{\sigma\sigma}^{\spec{3}{J}_J,\spec{3}{J}_J} &0&0
		\\[0.5ex]
		V_{\pi\d}^{\spec{3}{(J-1)}_J,\spec{3}{J}_J} &0&0&0
		\\[0.5ex]
		V_{\pi\d}^{\spec{3}{(J+1)}_J,\spec{3}{J}_J} &0&0&0
	\end{pmatrix}\,,
	\\
	\underline{\underline{\mathcal{G}}}^{(\text{NLO})\,[J,J]}
	=&\ \text{diag}\left(
		\mathcal{G}_\d^{\text{(NLO)}},\,\mathcal{G}_\sigma^{\text{(NLO)}},\,\mathcal{G}_\pi^{\text{(LO)}},\,\mathcal{G}_\pi^{\text{(LO)}}
	\right),
\end{align}
and
\begin{align}
	\underline{\underline{K}}^{\text{(NLO)}\,[J\pm1,J]}
	=&\ \begin{pmatrix}
		\Gamma_{\d\d}^{\spec{3}{(J\pm1)}_J,\spec{3}{(J\pm1)}_J} & \!\!\!\!(V_{\d\sigma}+\Gamma_{\d\sigma})^{\spec{3}{(J\pm1)}_J,\spec{3}{(J\pm1)}_J} & V_{\d\pi}^{\spec{3}{(J\pm1)}_J,\spec{\overline{2\mp1}}{J}_J}
		\\[0.5ex]
		(V_{\sigma\d}+\Gamma_{\sigma\d})^{\spec{3}{(J\pm1)}_J,\spec{3}{(J\pm1)}_J} & \Gamma_{\sigma\sigma}^{\spec{3}{(J\pm1)}_J,\spec{3}{(J\pm1)}_J} &0
		\\[0.5ex]
		V_{\pi\d}^{\spec{\overline{2\mp1}}{J}_J,\spec{3}{(J\pm1)}_J} &0&0
	\end{pmatrix}\,,
	\\
	\underline{\underline{\mathcal{G}}}^{(\text{NLO})\,[J\pm1,J]}
	=&\ \text{diag}\left(
		\mathcal{G}_\d^{\text{(NLO)}},\,\mathcal{G}_\sigma^{\text{(NLO)}},\,\mathcal{G}_\pi^{\text{(LO)}}
	\right),
\end{align}
respectively, similar to Eq.~\eqref{Eq:LSEq_PW}. The propagator function
$\mathcal{G}_\pi^\text{(LO)}$ is defined via
Eq.~\eqref{Eq:NewPropFunctions} with $a=\pi$ and reduced mass
\mbox{$\mu_\pi=\mu_\sigma=\mN(\mN+\mc)/(2\mN+\mc)$}.

\bibliography{HaloReactions.bbl}
\end{document}